\documentclass{article}%
\usepackage{amsmath}
\usepackage{graphicx}
\usepackage{amsfonts}
\usepackage{amssymb}%
\providecommand{\U}[1]{\protect\rule{.1in}{.1in}}

\begin{document}

\title{Theory of matter wave beam splitters in gravito-inertial and trapping potentials}
\author{Charles Antoine$^{1,2}$ and Christian J. Bord\'{e}$^{3,4}$\\$^{1}${\small \textit{ERGA, LERMA, CNRS-Observatoire de Paris,}}\\{\small \textit{Universit\'{e} P. et M. Curie, 75005 Paris, France}}\\$^{2}${\small \textit{Group of Gravitation and Experiments with Cold Atoms,
TIFR, 400005\ Mumbai, India}}\\$^{3}${\small \textit{BNM-SYRTE, CNRS-Observatoire de Paris, 75014 Paris,
France}}\\$^{4}${\small \textit{LPL, CNRS, Universit\'{e} Paris Nord, 93430
Villetaneuse, France}}}
\date{July 16, 2005}
\maketitle

\begin{abstract}
We present a strong field theory of matter wave splitting in the presence of
various gravitational, inertial and trapping potentials. The effect of these
potentials on the resonance condition (between the splitting potential and the
considered effective two-level system) and on the atomic Borrmann effect is
investigated in detail. The dispersive structuring of an incident atomic wave
packet - due to such generalized beam splitters - is studied and modeled, and
several important dynamical features of the solutions are detailed
(generalized Rabi oscillations, velocity selection, anomalous dispersion,
generalized Borrmann effect and anomalous gravitational bending). Finally, we
show how to express this triple interaction \textquotedblleft matter -
splitting potential - gravito-inertial and trapping
potentials\textquotedblright\ as an equivalent instantaneous interaction which
turns out to be a very efficient tool for the modeling of atom interferometers.

\bigskip\noindent PACS number(s): 03.75.-b, 32.80.-t, 33.80.-b, 39.25.+k, 42.50.-p

\end{abstract}

\section{Introduction}

\bigskip\ 

The matter wave beam splitters are nowadays the cornerstone of a wide range of
experiments, from atomic clocks and gravito-inertial sensors to laser cooling
and ultracold atoms characterization, quantum computing and cavity QED
experiments, atom lithography and chemical reaction dynamics, detection of
tiny effects of General Relativity and test of fundamental theories,
measurement of atom surface interactions...

In view of the recent progress in non-dissipative atom optics (coherent beam
splitters, mirrors, lenses...) as well as in dissipative atom optics (slowing,
trapping and cooling of atoms and molecules), it is needed to deepen our
comprehension of light-matter interactions in the presence of other external
potentials, like gravito-inertial or trapping potentials. In particular, the
precision and stability of atom interferometers are now so outstanding
\cite{gustavson00} that it is necessary to go beyond the former modeling of
their main component, namely the beam splitters.

In fact, the concept of atomic beam splitter is not confined to light-matter
interactions, and can be extended to any interaction process between matter
waves. It is thus possible to write the action of such atom optical elements
as a S matrix between the incident and diffused matter waves, where the S
matrix depends mainly on the \textquotedblleft splitting\textquotedblright%
\ potential, which can be material (slits, periodic microstructures...) or
electromagnetic (magnetic or electric static fields, laser fields...). This S
matrix description is particularly useful for the modeling of atom
interferometers, and more generally for any set up having a succession of such
beam splitters \cite{borde94,antoinejopb,antoineThese}.

However, for a long time, the precision of atom optics experiments has
remained low enough not to require an accurate study of matter wave beam
splitters. Thus, in the most common simplified modeling of these elements,
only the following effects were considered: 1) the splitting of an incident
atomic wave packet into several wave packets, 2) among them one was equal to
the incident wave packet, up to a change of amplitude, 3) and where the others
could differ from the incident wave packet in their central momentum, internal
state, amplitude and phase. However, this practical modeling - sometimes
called \textquotedblleft infinitely thin\textquotedblright\ because it amounts
to neglecting the duration of the interaction - does not take into account
several important effects, like the dispersive structuring of the incident
wave packet (velocity selection and sidebands, Borrmann effect, anomalous
dispersion...), the time and space dependency of the splitting potential, the
effect of relaxation processes, or the effect of other external fields during
the splitting (time-dependent gravito-inertial effects, trapping potentials...).

During the last past two decades, several authors studied some of these
problems, namely: the effect of a non-trivial time dependency of the splitting
potential (for running and standing laser beam splitters)
\cite{hioe85,suominen92,ishikawa94,carmel00,ishkhanyan00}; the atomic Borrmann
effect and anomalous dispersion effect without any other external potential
\cite{oberthaler96,eiermann03}\ or with a constant and uniform acceleration
(WKB solution \cite{lammerzahl99}); the atomic splitting in a constant and
uniform acceleration (exact solution in the temporal case and WKB solution in
the spatial case) \cite{lammerzahl95}; a common modeling for both spatial and
temporal beam splitters to the first order in the splitting potential (weak
field theory) \cite{borde04}...

In the light of what happened in neutron optics, where the beam splitters
modeling proved to be crucial to understand properly the origin of the
interferometer phase shifts \cite{horne86,rauch00}, it appears to be necessary
to go beyond these studies, so as to provide a comprehensive modeling of the
true action of a matter wave beam splitter (strong fields theory for all the
involved external fields).

This paper is organized as follows. First, we give some details on our
framework and explain how to put in equation the problem of the triple
interaction \textquotedblleft matter - splitting potential - other external
fields\textquotedblright. Then, we detail how to transform the obtained
equation in a simpler one thanks to unitary transformations (interaction
picture) and passage into the rotating frames. We expound then how to solve
this equation (analytically or numerically, with different developments or
relevant approximations), and we go back to the initial representation to
explain how to write the effect of such beam splitters as an effective
instantaneous interaction (generalized $ttt$ scheme). Finally, we study the
atomic Borrmann effect and other anomalous dispersive properties and model
them in the general framework detailed in the second part.

\bigskip\ 

\section{Framework and approximations\label{part2}}

\bigskip

\subsection{General framework}

\bigskip

The matter wave beam splitters we consider in this paper consist of
multi-level atomic systems subject to an interaction potential which couples
the levels together. This interaction is usually made in the presence of other
external potentials and miscellaneous relaxation processes.

In fact, these \textquotedblleft atomic systems\textquotedblright\ can refer
to atoms (neutral or not) as well as molecules, and more generally to any
quantity of matter which can be coherently manipulated.

Furthermore, the atomic levels are not restricted to internal atomic levels,
but more generally refer to energy-momentum states (i.e. eigenstates of both
the internal and kinetic Hamiltonians). The transitions can occur between
internal states only (spectroscopy without Doppler effect for example),
external states only (diffraction in Kapitza-Dirac and Bragg regimes, optical
Stern-Gerlach effect, magnetic atom mirror...), or between entangled states,
where the entanglement may be between the internal and external states
(stimulated Raman transitions for example) or between the previous
energy-momentum states and the eigenstates of the interaction potential (Fock
states of the quantized electromagnetic field for example).

\bigskip

There are many techniques to coherently split a matter wave, and each of them
corresponds to a particular kind of splitting potential. Like the other atom
optics elements, matter wave beam splitters use essentially two properties of
atoms (or molecules): their wave property, and their interaction with external
fields, electromagnetic or material. To date, the demonstrated matter wave
beam splitters are based on:

\begin{enumerate}
\item atom-matter interaction: front wave division (material slits)
\cite{carnal91}, amplitude splitting (transmission material gratings)
\cite{keith88}, reflection at crystalline surface \cite{clauser88}, quantum
reflection \cite{shimizu01}\ldots

\item interaction with static magnetic or electric fields (for atoms having a
dipolar electric or magnetic moment): longitudinal Stark effect
\cite{sokolov73}, transverse and longitudinal Stern-Gerlach effects
\cite{robert91}, magnetic mirrors \cite{opat92}, mirror for polar molecules
\cite{wark92}, Y shaped magnetic guides \cite{cassettari00}\ldots

\item resonant or quasi-resonant interaction with laser fields: reflection and
diffraction by standing \cite{arimondo79}, running
\cite{baklanov76,kasevich91} or evanescent \cite{cook82} laser waves (with
spatial and/or temporal working), optical Stern-Gerlach effect
\cite{kazantsev75}, stimulated Raman effect \cite{kasevich91} and its
derivatives (adiabatic transfer \cite{oreg84}, STIRAP \cite{gaubatz88}, CHIRAP
\cite{band93}\ldots), magneto-optical beam splitters \cite{pfau93}, X shaped
dipolar guides \cite{houde00}\ldots
\end{enumerate}

In this paper, we will focus on this third kind of interaction, and more
generally on the beam splitters for which the two-beam approximation is valid.

As for the relaxation processes, they refer to all the processes which lead to
a loss of coherence and/or a loss of atoms (spontaneous emission, inter-atomic
collisions, absorption and interaction with the material microstructures\ldots
). When they can not be neglected, the use of a density operator formalism is needed.

Other external potentials may be present during the matter wave splitting:
inertial and gravitational fields, trapping potential, van der Waals and
Casimir potentials\ldots\ In this paper, we consider all the time-dependent
potentials which are at most quadratic in position and momentum. The
corresponding Hamiltonian is therefore:%

\[
H_{ext}=\frac{1}{2m}\overrightarrow{p}.\overset{\Rightarrow}{\beta}\left(
t\right)  .\overrightarrow{p}-\frac{m}{2}\overrightarrow{r}.\overset
{\Rightarrow}{\gamma}\left(  t\right)  .\overrightarrow{r}-\overrightarrow
{r}.\overset{\Rightarrow}{\alpha}\left(  t\right)  .\overrightarrow
{p}-m\overrightarrow{g}\left(  t\right)  .\overrightarrow{r}+\overrightarrow
{f}\left(  t\right)  .\overrightarrow{p}%
\]

This includes the effect of non-uniform accelerations ($\overrightarrow
{g}\left(  t\right)  $ and $\overset{\Rightarrow}{\gamma}\left(  t\right)  $),
rotations (with an angular velocity $\overrightarrow{\Omega}\left(  t\right)
$ such that $\overset{\Rightarrow}{\alpha}\left(  t\right)  .\overrightarrow
{u}:=-\overrightarrow{\Omega}\left(  t\right)  \times\overrightarrow{u}$ for
any vector $\overrightarrow{u}$), trapping potentials ($-$ $\overset
{\Rightarrow}{\gamma}\left(  t\right)  $), non-zero curvature tensor
($\overset{\Rightarrow}{\gamma}\left(  t\right)  $), gravitational waves in
Fermi's gauge ($\overset{\Rightarrow}{\gamma}\left(  t\right)  $) or
Einstein's gauge ($\overset{\Rightarrow}{\beta}\left(  t\right)  $), and all
the electromagnetic potentials which can be written as a development at most
quadratic in position and momentum ($\overrightarrow{g}\left(  t\right)  $,
$\overset{\Rightarrow}{\gamma}\left(  t\right)  $, $\overrightarrow{f}\left(
t\right)  $). Furthermore, to keep an overall approach, the coefficients of
$H_{ext}$ are time-dependent, and $\overset{\Rightarrow}{\alpha}$,
$\overset{\Rightarrow}{\beta}$ and $\overset{\Rightarrow}{\gamma}$ are
expressed with non-diagonal 3x3 matrices.

\bigskip

\subsection{Approximations considered in this paper}

\bigskip

It is often possible to simplify this general framework and obtain an
evolution equation between only two effective states by using some justified approximations.

First, the two-beam approximation is indeed valid when only two
energy-momentum eigenstates are coupled. The coupling can be direct (for true
two-level systems) or indirect (Raman transitions, spatial beam splitters in
Bragg regime). In fact, one can show that any N-photon transition (thanks to
several running or standing laser waves) of a multilevel atom is equivalent to
an effective 1-photon transition between two atomic levels if the other levels
can be adiabatically eliminated \cite{shore91,borde97}. The effective photon
may not be real. For example, in the Bragg regime, the wave vector of this
effective photon is equal to $N\hbar\overrightarrow{k}$ and its frequency is
equal to $0$. Eventually, the spatial and temporal structure of the true laser
beams appears only in the amplitude of the effective running laser beam.

Second, the laser fields are considered as classical (coherent states of the
quantized electromagnetic fields), but the calculations which follow are also
valid for a transition between two dressed states \cite{cohen92}.

Third, we suppose that the two atomic levels have a long lifetime and we
neglect all the relaxation processes listed before. However, the instability
of these levels, due to spontaneous emission, can be taken into account in an
approximative manner adding a non-Hermitian part to the atom-laser Hamitonian
$V_{em}$ \cite{cohen92}.

In addition, $V_{em}$ is chosen equal to the usual dipolar electric
Hamiltonian (without spin) and we suppose that the other external fields are
sufficiently weak to neglect their effect on the atomic levels and laser fields.

Finally, the triple interaction \textquotedblleft laser - matter - other
external fields\textquotedblright\ can be written as a Schr\"{o}dinger
equation concerning two atomic states coupled by an effective running laser wave:%

\begin{equation}
i\hbar\frac{d}{dt}\left|  \Psi\left(  t\right)  \right\rangle =\left(
H_{0}+H_{ext}\left(  \overrightarrow{r_{op}},\overrightarrow{p_{op}},t\right)
+V_{em}\left(  \overrightarrow{r_{op}},t\right)  \right)  \left|  \Psi\left(
t\right)  \right\rangle \label{eq1}%
\end{equation}

where $\overrightarrow{r_{op}}$ and $\overrightarrow{r_{op}}$ are the position
and momentum operators, and $H_{0}$ the internal Hamiltonian ($E_{b}>E_{a}$):%

\[
H_{0}=\left(
\begin{array}
[c]{cc}%
E_{b} & 0\\
0 & E_{a}%
\end{array}
\right)  =\frac{E_{a}+E_{b}}{2}1+\frac{\hbar\omega_{0}}{2}\sigma_{3}%
\]

where $\omega_{0}=\left(  E_{b}-E_{a}\right)  /\hbar$ is the atomic transition
frequency and $\sigma_{3}$\ the usual third Pauli matrix.

It is also possible to account for some relativistic effects by introducing
two different masses \cite{borde04,antoineThese}. For simplicity however, we
will not take into account these relativitic corrections and use only one
atomic mass in what follows.

\bigskip

\section{Interaction picture and rotating frames\label{part3}}

\bigskip

It is generally impossible to solve directly the equation (\ref{eq1})
(non-trivial time dependence of the right hand side, presence of two
non-commuting operators $\overrightarrow{r_{op}}$ and $\overrightarrow{r_{op}%
}$), but it is possible to simplify it with the help of well chosen unitary
transformations \cite{antoineThese}.

The main idea of this series of transformations is to eliminate progressively
the different sources of evolution (internal and external) of the right hand
side of (\ref{eq1}). As each unitary transformation corresponds to a change of
frame, we can see this succession of transformations as a succession of frame
changes which aims at reaching the proper frame of the atom, or, at least, at
reaching a \textquotedblleft least movement frame\textquotedblright\ for the
atom (from external as well as internal points of view). In this especially
suitable frame, it is easier to solve the evolution equation and several
important pieces of information about the solution can be directly seen.

\bigskip

First, let us go to the interaction picture with respect to $H_{0}$
and$\ H_{ext}$:%

\begin{equation}
\left|  \Psi\left(  t\right)  \right\rangle =U_{1}\left(  t,t_{1}\right)
\left|  \varphi_{1}\left(  t\right)  \right\rangle \label{eqU1}%
\end{equation}

with :%
\[
U_{1}\left(  t,t_{1}\right)  =e^{-\frac{i}{\hbar}H_{0}.\left(  t-t_{1}\right)
}U_{ext}\left(  t,t_{1}\right)
\]

and:%

\[
U_{ext}\left(  t,t_{1}\right)  =\mathcal{T}\left(  \exp\left(  -\frac{i}%
{\hbar}\int_{t_{1}}^{t}H_{ext}\left(  \overrightarrow{r_{op}},\overrightarrow
{p_{op}},t^{\prime}\right)  dt^{\prime}\right)  \right)
\]

where $\mathcal{T}$ \ is the chronological Dyson operator and $t_{1}$ an
arbitrary time (different from $t$ by definition).

The equation (\ref{eq1}) becomes:%

\begin{equation}
i\hbar\frac{d}{dt}\left|  \varphi_{1}\left(  t\right)  \right\rangle
=e^{\frac{i}{\hbar}H_{0}.\left(  t-t_{1}\right)  }V_{em}\left(
\overrightarrow{R_{op}}\left(  t,t_{1}\right)  ,t\right)  e^{-\frac{i}{\hbar
}H_{0}.\left(  t-t_{1}\right)  }\left|  \varphi_{1}\left(  t\right)
\right\rangle \label{eq2}%
\end{equation}

where $\overrightarrow{R_{op}}$ is defined as:%

\[
\overrightarrow{R_{op}}\left(  t,t_{1}\right)  =U_{ext}\left(  t,t_{1}\right)
^{-1}\text{ }\overrightarrow{r_{op}}\text{ }U_{ext}\left(  t,t_{1}\right)
\]

The external Hamiltonian $H_{ext}$ being at most quadratic in position and
momentum, $\overrightarrow{R_{op}}$ depends linearly on $\overrightarrow
{r_{op}}$ and $\overrightarrow{p_{op}}$:%

\[
\overrightarrow{R_{op}}\left(  t,t_{1}\right)  =A\left(  t,t_{1}\right)
\text{ }\overrightarrow{r_{op}}+B\left(  t,t_{1}\right)  \text{ }%
\overrightarrow{p_{op}}/m+\overrightarrow{\xi}\left(  t,t_{1}\right)
\]

and is simply obtained through the classical solution of the Hamilton's
equations. One can show that the matrices $A$ and $B$ depend on the quadratic
terms of $H_{ext}$ ($\overset{\Rightarrow}{\alpha}\left(  t\right)  $,
$\overset{\Rightarrow}{\beta}\left(  t\right)  $ and $\overset{\Rightarrow
}{\gamma}\left(  t\right)  $) only, whereas $\overrightarrow{\xi}\left(
t,t_{1}\right)  $ also depends on its linear terms ($\overrightarrow{g}\left(
t\right)  $ and $\overrightarrow{f}\left(  t\right)  $). These matrices are in
fact the well known $ABCD$ matrices, usually used in Gaussian optics, and
introduced recently in atom optics \cite{borde91,antoineThese}.

\bigskip

As far as $V_{em}$ is concerned, it may have diagonal terms (AC stark shifts,
slowly varying in space and time). However, in a first approach, we can take
the latter constant and eliminate their common part by a unitary
transformation. Finally, $V_{em}$ can be taken as purely anti-diagonal:%

\[
V_{em}\left(  \overrightarrow{r_{op}},t\right)  =V\left(  \overrightarrow
{r_{op}},t\right)  \left(
\begin{array}
[c]{cc}%
0 & 1\\
1 & 0
\end{array}
\right)
\]

with:%

\[
V\left(  \overrightarrow{r},t\right)  =-2\hbar\Omega_{0}F\left(
\overrightarrow{r},t\right)  \cos\left(  \omega t-\overrightarrow
{k}.\overrightarrow{r}+\phi\right)
\]

where $F\left(  \overrightarrow{r},t\right)  $ is the amplitude of the
effective running laser beam, and where $\Omega_{0}$ is the Rabi frequency of
the atomic transition.

\bigskip

An other important approximation is the rotating wave approximation (RWA),
which consists in neglecting the off-resonant terms (i.e. with frequency
$\omega+\omega_{0}$) in (\ref{eq2}). In our case, this approximation is
supposed to be justified. Indeed, if the two effective levels are indirectly
coupled through optical photons (as for Raman transitions for example), one
can show that the Bloch-Siegert effect is negligible \cite{kasevich92bis}.

Finally, the evolution equation is equal to:%

\[
\frac{d}{dt}\left|  \varphi_{1}\left(  t\right)  \right\rangle =i\Omega
_{0}F\left(  \overrightarrow{R_{op}}\left(  t,t_{1}\right)  ,t\right)  \left(
\begin{array}
[c]{cc}%
0 & e^{-i\Phi_{op}\left(  t,t_{1}\right)  }\\
e^{+i\Phi_{op}\left(  t,t_{1}\right)  } & 0
\end{array}
\right)  \left|  \varphi_{1}\left(  t\right)  \right\rangle
\]

where $\Phi_{op}$\ is defined as:%

\[
\Phi_{op}\left(  t,t_{1}\right)  =\omega t-\omega_{0}\left(  t-t_{1}\right)
-\overrightarrow{k}.\overrightarrow{R_{op}}\left(  t,t_{1}\right)  +\phi
\]

\bigskip

The next unitary transformation corresponds to the usual passage into the
rotating frame. In fact, there is an infinity of such transformations
\cite{antoineThese}. For example, the following family of transformations
(indexed by the real parameter $x$):%

\begin{equation}
\left|  \varphi_{1}\left(  t\right)  \right\rangle =U_{2}\left(
x,t,t_{1}\right)  \left|  \varphi_{2}\left(  x,t\right)  \right\rangle
\label{eqU2}%
\end{equation}

with:%

\[
U_{2}\left(  x,t,t_{1}\right)  =\left(
\begin{array}
[c]{cc}%
e^{-i\Phi_{op}\left(  t,t_{1}\right)  .x} & 0\\
0 & e^{-i\Phi_{op}\left(  t,t_{1}\right)  .\left(  x-1\right)  }%
\end{array}
\right)
\]

leads to the equation:%

\begin{equation}
\fbox{$\frac{d}{dt}\left\vert \varphi_{2}\left(  x,t\right)  \right\rangle
=iM_{op}\left(  x,t\right)  \left\vert \varphi_{2}\left(  x,t\right)
\right\rangle $} \label{eq3}%
\end{equation}

with $M_{op}$\ defined as:%

\[
M_{op}\left(  x,t\right)  =\left(
\begin{array}
[c]{cc}%
x\left(  \omega-\omega_{0}-\overrightarrow{k}.\overset{\cdot}{\overrightarrow
{R_{op}}}-x\delta\right)  & \Omega_{0}F_{op}\\
\Omega_{0}F_{op} & \left(  x-1\right)  \left(  \omega-\omega_{0}%
-\overrightarrow{k}.\overset{\cdot}{\overrightarrow{R_{op}}}-\left(
x-1\right)  \delta\right)
\end{array}
\right)
\]

where $\delta\left(  t,t_{1}\right)  $ is the ``generalized recoil'':%

\[
\delta\left(  t,t_{1}\right)  =\hbar\overrightarrow{k}\beta\overrightarrow
{k}/2m
\]
and where the point above letters refers to the time derivative.

Each value of $x$ corresponds to a particular evolution equation. For example,
the most ``symmetric'' choice is $x=1/2$, whereas the most ``physical'' choice
is $x=1$, and leads to:%

\[
M_{op}\left(  1,t\right)  =\left(
\begin{array}
[c]{cc}%
\Delta_{op1} & \Omega_{0}F_{op}\\
\Omega_{0}F_{op} & 0
\end{array}
\right)
\]

where $\Delta_{op1}$ is the ``generalized detuning'' \cite{antoineThese}:%

\[
\Delta_{op1}\left(  t,t_{1}\right)  =\omega-\omega_{0}-\overrightarrow
{k}.\overset{\cdot}{\overrightarrow{R_{op}}}-\delta
\]

i.e., the operator which generalizes the usual ``free'' detuning in the
presence of several gravitational, inertial and trapping potentials. It can
also be written as:%

\[
\Delta_{op1}\left(  t,t_{1}\right)  =\omega-\left(  \omega_{0}+\overrightarrow
{k}\left[  \frac{\beta}{m}\left(  \overrightarrow{P_{op}}+\frac{\hbar
\overrightarrow{k}}{2}\right)  +\alpha\overrightarrow{R_{op}}+\overrightarrow
{f}\right]  \right)
\]

where $\overrightarrow{P_{op}}$ is defined in the same way as $\overrightarrow
{R_{op}}$ \cite{antoinejopb,antoineThese}:%

\[
\overrightarrow{P_{op}}\left(  t,t_{1}\right)  /m=C\left(  t,t_{1}\right)
\text{ }\overrightarrow{r_{op}}+D\left(  t,t_{1}\right)  \text{ }%
\overrightarrow{p_{op}}/m+\overrightarrow{\phi}\left(  t,t_{1}\right)
\]

\bigskip

The expression of $\Delta_{op1}\left(  t,t_{1}\right)  $ can be easily
interpreted considering the energy-momentum conservation for a non-excited
atom absorbing a photon ($\omega$, $\overrightarrow{k}$) at the instant $t$
(the considered atom is then at the position $\overrightarrow{R}\left(
t,t_{0}\right)  $ with the momentum $\overrightarrow{P}\left(  t,t_{0}\right)
$ when the absorption occurs):%

\[
H_{ext}\left(  \overrightarrow{R}\left(  t,t_{0}\right)  ,\overrightarrow
{P}\left(  t,t_{0}\right)  ,t\right)  +E_{b}+\hbar\omega\text{ }=\text{
}H_{ext}\left(  \overrightarrow{R}\left(  t,t_{0}\right)  ,\overrightarrow
{P}\left(  t,t_{0}\right)  +\hbar\overrightarrow{k},t\right)  +E_{a}%
\]

which gives the non-operatorial version of the condition $\Delta_{op1}\left(
t,t_{1}\right)  =0$ (exact resonance condition in the presence of the external
potentials described by $H_{ext}$).

This generalized detuning can also be expressed directly with the coefficients
of $H_{ext}$. For example, if $H_{ext}$ is constant, the first terms of its
Taylor expansion (in $\alpha(t-t_{1})$ and $\gamma(t-t_{1})^{2}$) are found to
be \cite{antoineThese}:%

\begin{align}
\Delta_{op1}\left(  t,t_{1}\right)   &  =\omega-\omega_{0}-\overrightarrow
{k}.\frac{\overrightarrow{p_{op}}}{m}-\delta-\overrightarrow{k}%
.\overrightarrow{g}\left(  t-t_{1}\right)  -\overrightarrow{k}.\alpha
.\overrightarrow{r_{op}}\label{eqdeltaop}\\
&  -2\overrightarrow{k}.\alpha.\frac{\overrightarrow{p_{op}}}{m}\left(
t-t_{1}\right)  -\overrightarrow{k}.\left(  \alpha^{2}+\gamma\right)
.\overrightarrow{r_{op}}\left(  t-t_{1}\right)  -\overrightarrow{k}%
.\alpha.\overrightarrow{g}\left(  t-t_{1}\right)  ^{2}-...\nonumber
\end{align}

where only the first five terms of the right hand side are non-negligible in
usual experiments (weak rotations and acceleration gradients on the Earth). We
can then use chirped laser pulses to eliminate the gravitational induced
Doppler shift $\overrightarrow{k}.\overrightarrow{g}\left(  t-t_{1}\right)  $
and finally get back the usual ``free'' detuning.

The other element of $M_{op}\left(  x,t\right)  $ which may depend on the two
canonical operators is the effective amplitude $F\left(  \overrightarrow
{R_{op}}\left(  t,t_{1}\right)  ,t\right)  $. The two main sources of its
spatial dependency are its transverse profile and the speckle due to the
miscellaneous optical elements used to bring the lasers to the atoms.
Generally, one can not neglect this speckle and the best is to map it, and
then, to use the zones where the speckle is sufficiently weak.

Concerning the laser transverse profile, one can show that it is seen roughly
uniform by each individual atom of the initial atomic cloud (described by a
statistical mixture of wave packets). One can therefore replace
$\overrightarrow{R_{op}}\left(  t,t_{1}\right)  $ in $F$ by its semi-classical
action on a typical wave packet which evolves inside such beam splitters. For
example, and as we see thereafter, $\overrightarrow{R_{op}}\left(
t,t_{1}\right)  $ can be approximated by:%

\[
\overrightarrow{R_{op}}\left(  t,t_{1}\right)  \simeq A\left(  t,t_{1}\right)
\text{ }\overrightarrow{r_{0}}+B\left(  t,t_{1}\right)  \left(  \text{
}\overrightarrow{p_{0}}+\hbar\overrightarrow{k}/2\right)  /m+\overrightarrow
{\xi}\left(  t,t_{1}\right)
\]

where $\overrightarrow{r_{0}}$ and $\overrightarrow{p_{0}}$ are the initial
central position and momentum of the considered atomic wave packet. The main
result is that $F\left(  \overrightarrow{R_{op}}\left(  t,t_{1}\right)
,t\right)  \simeq\overline{F}\left(  t,t_{1}\right)  $ is hence now
independent of $\overrightarrow{r_{op}}$ and $\overrightarrow{p_{op}}$ (but is
still time-dependent).

\bigskip

\section{Resolution methods\label{part4}}

\bigskip

The main problem in the integration of (\ref{eq3}) is that, in the general
case considered in this paper, $M_{op}\left(  x,t\right)  $ is a (2x2) matrix
which depends both on time and on the two non-commuting canonical operators
$\overrightarrow{r_{op}}$ and $\overrightarrow{p_{op}}$. These are the two
reasons why $M_{op}\left(  x,t\right)  $ does not commute with itself at
different times, and why one can not apply the common rules to integrate
(\ref{eq3}) directly.

However, in some particular cases, (\ref{eq3}) may depend on one
time-independent operator only, and one can solve it analytically in the
representation of this operator. It is thus important to list the maximum of
these exactly solvable cases. This is the aim of the $z(t)$ theory, initiated
in \cite{rosen32}, improved in \cite{demkov69,hioe85} and generalized recently
in \cite{ishkhanyan00} (for a detailed review, see \cite{ishkhanyan00}\ and
\cite{antoineThese}). Among these exact solutions, let us underline the
Landau-Zener model (solution with cylinder parabolic functions) which accounts
for the effect of a time-independent and uniform acceleration during the
atomic splitting, and the Rosen-Zener model (solution with hyperbolic secant
functions) which is significant in the study of matter wave solitons.

Some other analytical methods are particularly useful to deal with the
equation (\ref{eq3}): Floquet theory \cite{autler55} for periodic
time-dependence and its generalizations (multi-periodic Floquet method
\cite{ho84}, (t,t') theory \cite{peskin93}\ldots); bands theory (i.e. use of
Bloch states) when one can not make the RWA \cite{letokhov78}; use of
quasi-probabilities (Wigner and Shirley representations) and phase space
functions (Q function of Hushimi and Kano, P distribution of Glauber and
Sudarshan) when some QED effects can not be neglected (for a recent review,
see \cite{schleich01})\ldots

Apart from these particular exactly solvable cases, it is always possible to
write the general solution of (\ref{eq3}) as a formal development. This
development may be linear (Dyson) or not (Magnus, Fer, Cayley\ldots), and may
preserve the unitarity (for a review of the recent advances concerning the
Magnus expansion, see \cite{iserles00,moan01,antoineThese}). However, due to
the entanglement of operators $\overrightarrow{r_{op}}$ and $\overrightarrow
{p_{op}}$ in the different terms of these developments, it is impossible to
choose any representation which leads to an analytical expression of the
solution. This problem can be solved either in eliminating one of the two
canonical operators directly in the equation (\ref{eq3}) (``operatorial
elimination method''), or in approximating the generalized detuning
$\Delta_{op1}\left(  t,t_{1}\right)  $, or finally in solving the equation numerically.

The operatorial elimination method, which is detailed in \cite{antoineThese},
leads to a double development, easily calculable but rather long, which is why
its use would be limited to the numerical domain. As for the approximations,
one has already underlined that the effect of rotations and acceleration
gradients may often be neglected in the generalized detuning (leading to a
trivial integration of (\ref{eq3})). If not, several tactics may be used
\cite{antoineThese}:

\begin{enumerate}
\item freezing $\Delta_{op1}\left(  t,t_{1}\right)  $ at a particular time
(mid-time for example) or taking a temporal average, and choosing the
resulting time-independent representation;

\item or replacing $\overrightarrow{r_{op}}$ and $\overrightarrow{p_{op}}$ by
their semi-classical value (WKB approximation);

\item or replacing only one of these operators ($\overrightarrow{r_{op}}$ for
example) by its action on the initial atomic wave packet (or on a wave packet
closer to the final solution)
\end{enumerate}

When the equation (\ref{eq3}) is made scalar, one can use either the previous
analytical methods, or the previous developments (and truncate them when they
converge, see \cite{miao00,moan01}), or some intermediate approaches which are
based on the eigenstates of the matrix $M_{op}\left(  x,t\right)  $
(super-adiabatic scheme \cite{berry90} or successive adiabatic states method
\cite{antoineThese}). The latter are extremely interesting because they
directly provide important information on the solution, like the true energies
of the system ``atom -- laser -- other external fields'' and the corresponding
group velocities.

Several numerical methods may also be implemented (for a recent review, see
\cite{lubich02}): Magnus expansion, median exponential rule, Runge-Kutta
method, Strang-Marchuk-Trotter method, Chebyshev or Lanczos
approximations\ldots

\bigskip

Finally, we obtain a solution which is more or less close to the exact
solution of (\ref{eq3}):%

\[
\left|  \varphi_{sol}\left(  t\right)  \right\rangle =S_{op}\left(
t,t_{0}\right)  \left|  \varphi_{2}\left(  t_{0}\right)  \right\rangle
\simeq\left|  \varphi_{2}\left(  t\right)  \right\rangle
\]

with the evolution operator:%

\[
S_{op}\left(  t,t_{0}\right)  \simeq\mathcal{T}\left(  \exp\left(
i\int_{t_{0}}^{t}M_{op}\left(  t^{\prime}\right)  dt^{\prime}\right)  \right)
\]

which can be written explicitely as a $S$ matrix between the initial and final
atomic states:%

\[
\left(
\begin{array}
[c]{c}%
\left|  b_{sol}\left(  t\right)  \right\rangle \\
\left|  a_{sol}\left(  t\right)  \right\rangle
\end{array}
\right)  =\left(
\begin{array}
[c]{cc}%
S_{bb,op}\left(  t,t_{0}\right)  & S_{ba,op}\left(  t,t_{0}\right) \\
S_{ab,op}\left(  t,t_{0}\right)  & S_{aa,op}\left(  t,t_{0}\right)
\end{array}
\right)  \left(
\begin{array}
[c]{c}%
\left|  b_{2}\left(  t_{0}\right)  \right\rangle \\
\left|  a_{2}\left(  t_{0}\right)  \right\rangle
\end{array}
\right)
\]

The resulting action of this $S$ matrix on the initial atomic wave packet is a
possible change of internal state (generalized Rabi oscillations) and a
structuring into several wave packets that can be quite different from the
initial one. In particular, the group velocities of the created wave packets
may be identical or not (Borrmann effect, see below) according to the value of
the generalized detuning.

\bigskip

\section{Return to the initial picture and equivalent instantaneous
interaction\label{part5}}

\bigskip

\subsection{Solution in the initial representation}

\bigskip

Once we have obtained the matrix $S_{op}\left(  t,t_{0}\right)  $, we can do
the reverse unitary transformations of (\ref{eqU1})\ and (\ref{eqU2}), and go
back to the initial representation. One obtains (written here for the
symmetric choice $x=1/2$):%

\begin{align}
\left|  \Psi_{sol}\left(  t\right)  \right\rangle  &  =e^{-i\int_{t_{0}}%
^{t}\delta\left(  t^{\prime},t_{1}\right)  dt^{\prime}/4}.U_{1}\left(
t,t_{1}\right)  U_{2}\left(  1/2,t,t_{1}\right) \label{eq4}\\
&  S_{op}\left(  t,t_{0}\right)  \text{ }U_{2}^{-1}\left(  1/2,t_{0}%
,t_{1}\right)  U_{1}^{-1}\left(  t_{0},t_{1}\right)  \left|  \Psi\left(
t_{0}\right)  \right\rangle \nonumber
\end{align}

which expresses the link between the general solution $\left|  \Psi
_{sol}\left(  t\right)  \right\rangle $ and the initial ket $\left|
\Psi\left(  t_{0}\right)  \right\rangle $.

It may be noticed that the solution depends on $t_{1}$, the arbitrary time
introduced to define the previous unitary transformations. This instant has no
physical meaning and can be removed explicitly at each step of our
calculations \cite{antoineThese}. However, it is more interesting to keep it
and eventually assign it a particular value (the central time of the
interaction for example) which may be useful both for calculations and the
interpretation of the obtained solutions. The relevance of this instant to
describe the atomic beam splitting as an equivalent instantaneous interaction
(generalized $ttt$ scheme) will appear more clearly at the end of this part.

It is easy to interpret the expression of $\left|  \Psi_{sol}\left(  t\right)
\right\rangle $ by writing the typical terms $\exp\left(  \pm\frac{i}{2}%
\Phi_{op}\left(  t,t_{0}\right)  \right)  $ of $U_{2}^{\pm1}$ as a product of
exponentials. Indeed, let us consider the following initial ket:%

\[
\left|  \Psi\left(  t_{0}\right)  \right\rangle =\left(
\begin{array}
[c]{c}%
0\\
\left|  a\left(  t_{0}\right)  \right\rangle
\end{array}
\right)
\]

which represents the incident atomic wave packet (atoms in the lower state
$a$). It results that the upper component of $\left|  \Psi_{sol}\left(
t\right)  \right\rangle $ is:%

\begin{align}
&  e^{i\theta_{1ba}}\text{ }U_{ext}\left(  t,t_{1}\right)  \text{ }e^{\frac
{i}{2}\overrightarrow{k}A\left(  t,t_{1}\right)  \overrightarrow{r_{op}}%
}\text{ }e^{\frac{i}{2}\overrightarrow{k}B\left(  t,t_{1}\right)
\overrightarrow{p_{op}}/m}\text{ }\label{eq5}\\
&  S_{ba,op}\left(  t,t_{0}\right)  \text{ }e^{\frac{i}{2}\overrightarrow
{k}A\left(  t_{0},t_{1}\right)  \overrightarrow{r_{op}}}\text{ }e^{\frac{i}%
{2}\overrightarrow{k}B\left(  t_{0},t_{1}\right)  \overrightarrow{p_{op}}%
/m}\text{ }U_{ext}\left(  t_{1},t_{0}\right)  \text{ }\left|  a\left(
t_{0}\right)  \right\rangle \nonumber
\end{align}

which can be interpreted as follows:

\begin{enumerate}
\item evolution from $t_{0}$ to $t_{1}$ due to $U_{ext}$, i.e. to the
gravito-inertial and trapping potentials described by $H_{ext}$

\item translation of $-\widetilde{B}\left(  t_{0},t_{1}\right)  \frac
{\hbar\overrightarrow{k}}{2m}$ in position and $+\widetilde{A}\left(
t_{0},t_{1}\right)  \frac{\hbar\overrightarrow{k}}{2}$ in momentum due to the
term $\exp\left(  \frac{i}{2}\overrightarrow{k}A\left(  t_{0},t_{1}\right)
\overrightarrow{r_{op}}\right)  .\exp\left(  \frac{i}{2}\overrightarrow
{k}B\left(  t_{0},t_{1}\right)  \overrightarrow{p_{op}}/m\right)  $

\item action of the non-diagonal element of the $S$ matrix $S_{ba,op}\left(
t,t_{0}\right)  $ (change of internal state and dispersive structuring)

\item again two translations: $-\widetilde{B}\left(  t,t_{1}\right)
\frac{\hbar\overrightarrow{k}}{2m}$ in position and $+\widetilde{A}\left(
t,t_{1}\right)  \frac{\hbar\overrightarrow{k}}{2}$ in momentum

\item then the external evolution from $t_{1}$ to $t$ due to $H_{ext}$

\item and finally a phase factor $\exp\left(  i\theta_{1ba}\right)  $ with
\end{enumerate}

\begin{align*}
\theta_{1ba}  &  =-\frac{E_{b}\left(  t-t_{0}\right)  }{\hbar}-\frac{\left(
\omega-\omega_{0}\right)  \left(  t-t_{0}\right)  -\overrightarrow{k}\left(
\overrightarrow{\xi}\left(  t,t_{1}\right)  +\overrightarrow{\xi}\left(
t_{0},t_{1}\right)  \right)  }{2}-\omega t_{0}-\phi\\
&  +\frac{\hbar\overrightarrow{k}}{8m}\left(  A\left(  t,t_{1}\right)
\widetilde{B}\left(  t,t_{1}\right)  +A\left(  t_{0},t_{1}\right)
\widetilde{B}\left(  t_{0},t_{1}\right)  \right)  \overrightarrow{k}%
-\int_{t_{0}}^{t}\frac{\delta\left(  t^{\prime},t_{1}\right)  }{4}dt^{\prime}%
\end{align*}

Similarly, the lower component of $\left|  \Psi_{sol}\left(  t\right)
\right\rangle $ is equal to:%

\begin{align}
&  e^{i\theta_{1aa}}\text{ }U_{ext}\left(  t,t_{1}\right)  \text{ }%
e^{-\frac{i}{2}\overrightarrow{k}A\left(  t,t_{1}\right)  \overrightarrow
{r_{op}}}e^{-\frac{i}{2}\overrightarrow{k}B\left(  t,t_{1}\right)
\overrightarrow{p_{op}}/m}\label{eq6}\\
&  \text{ }S_{aa,op}\text{ }e^{\frac{i}{2}\overrightarrow{k}A\left(
t_{0},t_{1}\right)  \overrightarrow{r_{op}}}e^{\frac{i}{2}\overrightarrow
{k}B\left(  t_{0},t_{1}\right)  \overrightarrow{p_{op}}/m}\text{ }%
U_{ext}\left(  t_{1},t_{0}\right)  \left|  a\left(  t_{0}\right)
\right\rangle \nonumber
\end{align}

with the following similar interpretation:

\begin{enumerate}
\item evolution from $t_{0}$ to $t_{1}$ due to $H_{ext}$

\item translations of $-\widetilde{B}\left(  t_{0},t_{1}\right)  \frac
{\hbar\overrightarrow{k}}{2m}$\textit{ }in position and of $+\widetilde
{A}\left(  t_{0},t_{1}\right)  \frac{\hbar\overrightarrow{k}}{2}$ in momentum

\item action of the diagonal element $S_{aa,op}\left(  t,t_{0}\right)  $ of
the $S$ matrix (dispersive structuring without internal change)

\item translations of $+\widetilde{B}\left(  t,t_{1}\right)  \frac
{\hbar\overrightarrow{k}}{2m}$\textit{\ } in position and of $-\widetilde
{A}\left(  t,t_{1}\right)  \frac{\hbar\overrightarrow{k}}{2}$ in momentum

\item evolution from $t_{1}$ to $t$ due to $H_{ext}$

\item and a phase factor $\exp\left(  i\theta_{1aa}\right)  $ with
\end{enumerate}

\begin{align*}
\theta_{1aa}  &  =-\frac{E_{a}\left(  t-t_{0}\right)  }{\hbar}+\frac{\left(
\omega-\omega_{0}\right)  \left(  t-t_{0}\right)  -\overrightarrow{k}\left(
\overrightarrow{\xi}\left(  t,t_{1}\right)  -\overrightarrow{\xi}\left(
t_{0},t_{1}\right)  \right)  }{2}\\
&  +\frac{\hbar\overrightarrow{k}}{8m}\left(  A\left(  t,t_{1}\right)
\widetilde{B}\left(  t,t_{1}\right)  +A\left(  t_{0},t_{1}\right)
\widetilde{B}\left(  t_{0},t_{1}\right)  \right)  \overrightarrow{k}%
-\int_{t_{0}}^{t}\frac{\delta\left(  t^{\prime},t_{1}\right)  }{4}dt^{\prime}%
\end{align*}

which differs from $\theta_{1ba}$ by $\omega t+\phi-\overrightarrow
{k}.\overrightarrow{\xi}\left(  t,t_{1}\right)  $.

\bigskip

\subsection{Generalized atomic Borrmann effect\label{part52}}

\bigskip

In certain conditions, the transfer matrix $S_{op}$ may have only a weak
effect on the external state of the incident atoms. In this case, the center
of the two main wave packets (associated with the upper and lower adiabatic
atomic states, see below) evolves along the same trajectory during the
splitting: it is the (atomic) Borrmann effect.

This effect is well known in dynamical diffraction of X rays \cite{borrmann41}%
, of neutron waves \cite{knowles56,rauch78}, and more recently of atomic waves
\cite{oberthaler96}. Historically, this effect was defined as a more specific
phenomenon, valid for any kind of waves diffracting in an absorbing crystal.
It can be stated as: ``for a certain angle of incidence with respect to the
crystal surface, the propagation of the incident wave packet inside the
crystal is made with no attenuation along a unique trajectory which is
orthogonal to the crystal surface''. This particular angle of incidence is the
well known Bragg angle defined as:%

\[
\overrightarrow{P}.\left(  \overrightarrow{p_{0}}+\overrightarrow{P}/2\right)
=0
\]

where $\overrightarrow{p_{0}}$ is the central momentum of the incident wave
packet and $\overrightarrow{P}$ is the quantum of momentum which is
communicated to the diffracted wave packet (the Bragg condition is written
here for the first order diffraction).

Conversely, two wave packets with two different trajectories are created
inside the crystal if this condition is not fulfilled (defining the well known
\textquotedblleft Borrmann fan\textquotedblright). Furthermore, this anomalous
transmission is very sensitive to the Bragg condition, and any deviation from
it greatly amplifies the angle between the two trajectories \cite{rauch00}.

In the case of atom-laser interactions, there is no absorption of atoms and
the Bragg condition (which means nothing but the energy-momentum conservation,
i.e. the resonance condition) is partly relaxed due to the atomic internal structure:%

\[
\omega-\omega_{0}-\overrightarrow{k}.\left(  \overrightarrow{p_{0}}%
+\hbar\overrightarrow{k}/2\right)  /m=0
\]

To date, this atomic Borrmann effect was only studied in the free case
\cite{oberthaler96} (for which $H_{ext}$ is limited to the usual kinetic
Hamiltonian $\overrightarrow{p}^{2}/2m$) or in the presence of a
time-independent and uniform acceleration \cite{lammerzahl99}. In this paper,
it is obtained in the presence of the various gravitational, inertial and
trapping potentials described by $H_{ext}$. By the way, we will show that the
common Borrmann trajectory is equal to the average of the two extreme
trajectories: $A$ $\overrightarrow{r_{0}}+B$ $\overrightarrow{p_{0}%
}/m+\overrightarrow{\xi}$\ (atom absorbing a photon at the final time $t$) and
$A$ $\overrightarrow{r_{0}}+B\left(  \text{ }\overrightarrow{p_{0}}%
+\hbar\overrightarrow{k}\right)  /m+\overrightarrow{\xi}$\ (atom absorbing a
photon at the initial time $t_{0}$).

Indeed, let us consider the previous example (atoms initially in state $a$)
and suppose that $S_{op}$ has a negligible effect on the central position and
momentum of the corresponding incident wave packet. Then, according to the
expression (\ref{eq4}) and the Ehrenfest theorem, we can show that the central
position $\overrightarrow{r_{0}}$ of the initial wave packet is changed into:%

\[
\overrightarrow{r_{c}}\left(  t,t_{0}\right)  =A\left(  \overrightarrow{r_{0}%
}-\widetilde{B}\frac{\hbar\overrightarrow{k}}{2m}\right)  +B\left(
\overrightarrow{p_{0}}+\left(  1+\widetilde{A}\right)  \frac{\hbar
\overrightarrow{k}}{2}\right)  /m+\overrightarrow{\xi}%
\]

where $\overrightarrow{p_{0}}$ is its initial central momentum.

Finally, thanks to some simple properties of $ABCD$ matrices, we obtain the
previously stated result:%

\begin{align*}
\overrightarrow{r_{c}}\left(  t,t_{0}\right)   &  =A\overrightarrow{r_{0}%
}+B\left(  \overrightarrow{p_{0}}+\frac{\hbar\overrightarrow{k}}{2}\right)
/m+\overrightarrow{\xi}\\
&  =\frac{1}{2}\left[  \left(  A\overrightarrow{r_{0}}+B\overrightarrow{p_{0}%
}/m+\overrightarrow{\xi}\right)  +\left(  A\overrightarrow{r_{0}}+B\left(
\overrightarrow{p_{0}}+\hbar\overrightarrow{k}\right)  /m+\overrightarrow{\xi
}\right)  \right]
\end{align*}

\begin{figure}
[h]
\begin{center}
\includegraphics[
trim=0.000000in 0.000000in -0.000627in 0.003758in,
width=352.3125pt
]%
{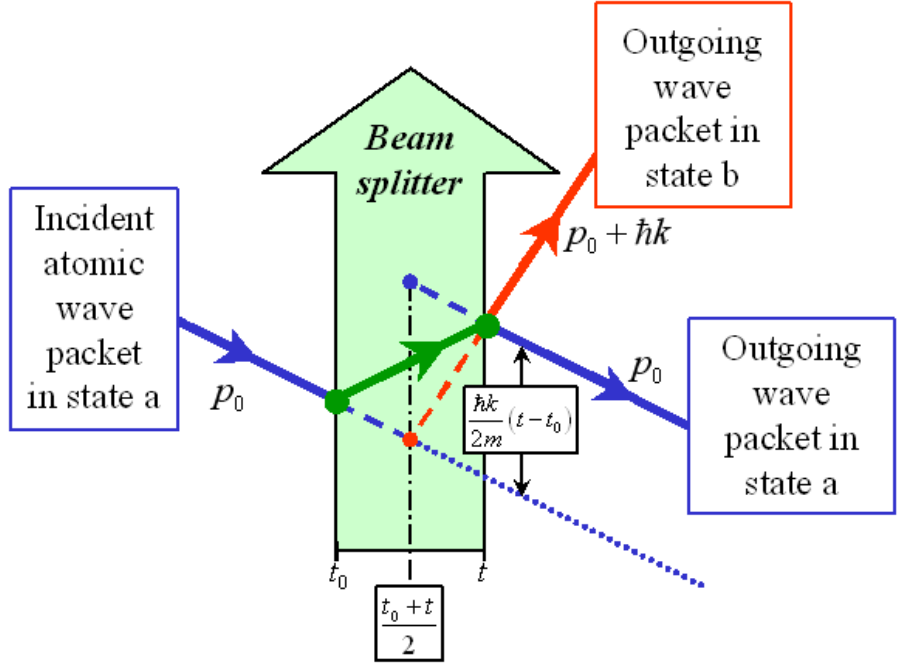}%
\caption{The continuous lines represent the central atomic trajectories (group
velocities); the dashed lines refer to the equivalent instantaneous
interaction (ttt scheme); and the dotted line is the extension of the initial
atomic trajectory (infinitely thin modeling).}%
\label{figborrmann}%
\end{center}
\end{figure}

This unique central trajectory differs by $B\left(  t,t_{0}\right)
.\hbar\overrightarrow{k}/2$\ from the one obtained without any splitting
potential. It means that, even for the atoms which are finally in the same
internal state as the initial one (no effective internal change), there is a
non-trivial change of their central trajectory, which results in a measurable
spatial shift at the end of the interaction (see Figure \ref{figborrmann}).%

\bigskip

As for the central momentum, we obtain similarly the two following momenta:%

\[
\overrightarrow{p_{c}}\left(  t,t_{0}\right)  =mC\overrightarrow{r_{0}%
}+D\left(  \overrightarrow{p_{0}}+\frac{\hbar\overrightarrow{k}}{2}\right)
+m\overrightarrow{\phi}\pm\frac{\hbar\overrightarrow{k}}{2}%
\]

which differ from each other on $\hbar\overrightarrow{k}$, as expected.

On the contrary, $S_{op}$ may have a non-negligible effect on the external
state of the incident wave packet (change of central position and momentum,
respectively due to the addition of a non-trivial group velocity and to the
change of momentum distribution). In this case, one can show that the initial
wave packet is split into two main wave packets, which evolve along two
different trajectories (i.e. with two distinct group velocities) which form
the atomic Borrmann fan. For each of these wave packets, we can do the same
calculation as before and obtain the trajectories in the initial frame
\cite{antoineThese}.

\bigskip

\subsection{Generalized ttt scheme\label{part53}}

\bigskip

However, in both previous cases, it is noticeable that the expression
(\ref{eq4}) can be written as a product of three evolution operators:%

\[
\left|  \Psi_{sol}\left(  t\right)  \right\rangle =U_{1}\left(  t,t_{1}%
\right)  \text{{\Large S}}_{_{1}}\left(  \overrightarrow{r_{op}}%
,\overrightarrow{p_{op}},t,t_{0},t_{1}\right)  U_{1}\left(  t_{1}%
,t_{0}\right)  \left|  \Psi\left(  t_{0}\right)  \right\rangle
\]

where $U_{1}\left(  t_{1},t_{0}\right)  $ and $U_{1}\left(  t,t_{1}\right)  $
describe the evolution due to $H_{0}+H_{ext}$ only, and where {\Large S}%
$_{_{1}}$ represents the evolution part depending on the splitting potential
$V$. The aim of this arrangement is to clearly separate the effect of
$H_{ext}$ and $V$, and to describe the interaction as an effective
instantaneous interaction (generalization of the $ttt$ scheme introduced in
\cite{borde04}).

In addition to the link with the infinitely thin modeling of atomic beam
splitters, we aim at providing a clear and practical beam splitter modeling,
in the presence of the external fields described by $H_{ext}$, which can be
used easily in atom interferometric phase shift calculations
\cite{antoinepla,antoinejopb,antoineThese}.

To clearly separate the splitting terms from the translation and phase terms
in (\ref{eq5}) and (\ref{eq6}), we can transform the following expression:%

\[
e^{\pm\frac{i}{2}\Phi_{op}\left(  t_{0},t_{1}\right)  }\text{ }S_{uv}\left(
\overrightarrow{r_{op}},\overrightarrow{p_{op}}\right)  \text{ }e^{\mp\frac
{i}{2}\Phi_{op}\left(  t_{0},t_{1}\right)  }%
\]

in \cite{antoineThese}:%

\[
S_{uv}\left(  \overrightarrow{r_{op}}\mp\widetilde{B}\left(  t_{0}%
,t_{1}\right)  \frac{\hbar\overrightarrow{k}}{2m},\overrightarrow{p_{op}}%
\pm\widetilde{A}\left(  t_{0},t_{1}\right)  \frac{\hbar\overrightarrow{k}}%
{2}\right)
\]

thanks to the following algebraic properties:%
\begin{align*}
e^{A}.f\left(  B\right)  .e^{-A}  &  =f\left(  e^{A}.B.e^{-A}\right) \\
e^{A}.B.e^{-A}  &  =B+\left[  A,B\right]  +\frac{1}{2!}\left[  A,\left[
A,B\right]  \right]  +...
\end{align*}
where $A$ and $B$ refer to operators or square matrices, and where $f$ is a function.

Finally, the diffusion matrix {\Large S}$_{1}$ can be written as:%

\[
\text{{\Large S}}_{1}=\left(
\begin{array}
[c]{cc}%
\text{{\Large S}}_{1,bb} & \text{{\Large S}}_{1,ba}\\
\text{{\Large S}}_{1,ab} & \text{{\Large S}}_{1,aa}%
\end{array}
\right)
\]

where its elements are equal to ($a$ and $b$ are the lower and upper states respectively):

\begin{enumerate}
\item for the $a\longrightarrow a$ transition:%
\[
\text{{\Large S}}_{1,aa}=e^{i\Phi_{aa}}e^{\frac{i}{\hbar}\overrightarrow
{r_{op}}.\overrightarrow{p_{aa}}}e^{-\frac{i}{\hbar}\overrightarrow{p_{op}%
}.\overrightarrow{r_{aa}}}S_{aa}\left(  \overrightarrow{r_{op}}-\widetilde
{B}\left(  t_{0},t_{1}\right)  \frac{\hbar\overrightarrow{k}}{2m}%
,\overrightarrow{p_{op}}+\widetilde{A}\left(  t_{0},t_{1}\right)  \frac
{\hbar\overrightarrow{k}}{2}\right)
\]
with:
\begin{align*}
\overrightarrow{p_{aa}}  &  =-\frac{\widetilde{A}\left(  t,t_{1}\right)
-\widetilde{A}\left(  t_{0},t_{1}\right)  }{2}\hbar\overrightarrow{k}\\
\overrightarrow{r_{aa}}  &  =+\frac{\widetilde{B}\left(  t,t_{1}\right)
-\widetilde{B}\left(  t_{0},t_{1}\right)  }{2}\frac{\hbar\overrightarrow{k}%
}{m}\\
\Phi_{aa}  &  =+\frac{1}{2}\left[  \left(  \omega-\omega_{0}\right)  \left(
t-t_{0}\right)  -\overrightarrow{k}.\left(  \overrightarrow{\xi}\left(
t,t_{1}\right)  -\overrightarrow{\xi}\left(  t_{0},t_{1}\right)  \right)
\right] \\
&  +\frac{\hbar\overrightarrow{k}}{8m}\left[  A\left(  t,t_{1}\right)
\widetilde{B}\left(  t,t_{1}\right)  +A\left(  t_{0},t_{1}\right)
\widetilde{B}\left(  t_{0},t_{1}\right)  -2A\left(  t_{0},t_{1}\right)
\widetilde{B}\left(  t,t_{1}\right)  \right]  \overrightarrow{k}-\int_{t_{0}%
}^{t}\frac{\delta\left(  t^{\prime},t_{1}\right)  }{4}dt^{\prime}%
\end{align*}

\item for the $a\longrightarrow b$ transition:%
\[
\text{{\Large S}}_{1,ba}=e^{i\Phi_{ba}}e^{\frac{i}{\hbar}\overrightarrow
{r_{op}}.\overrightarrow{p_{ba}}}e^{-\frac{i}{\hbar}\overrightarrow{p_{op}%
}.\overrightarrow{r_{ba}}}S_{ba}\left(  \overrightarrow{r_{op}}-\widetilde
{B}\left(  t_{0},t_{1}\right)  \frac{\hbar\overrightarrow{k}}{2m}%
,\overrightarrow{p_{op}}+\widetilde{A}\left(  t_{0},t_{1}\right)  \frac
{\hbar\overrightarrow{k}}{2}\right)
\]
with:
\begin{align*}
\overrightarrow{p_{ba}}  &  =+\frac{\widetilde{A}\left(  t,t_{1}\right)
+\widetilde{A}\left(  t_{0},t_{1}\right)  }{2}\hbar\overrightarrow{k}\\
\overrightarrow{r_{ba}}  &  =-\frac{\widetilde{B}\left(  t,t_{1}\right)
+\widetilde{B}\left(  t_{0},t_{1}\right)  }{2}\frac{\hbar\overrightarrow{k}%
}{m}\\
\Phi_{ba}  &  =-\left[  \omega\frac{t+t_{0}}{2}-\omega_{0}\left(
\frac{t+t_{0}}{2}-t_{1}\right)  +\phi-\overrightarrow{k}.\frac{\overrightarrow
{\xi}\left(  t,t_{1}\right)  +\overrightarrow{\xi}\left(  t_{0},t_{1}\right)
}{2}\right] \\
&  +\frac{\hbar\overrightarrow{k}}{8m}\left[  A\left(  t,t_{1}\right)
\widetilde{B}\left(  t,t_{1}\right)  +A\left(  t_{0},t_{1}\right)
\widetilde{B}\left(  t_{0},t_{1}\right)  +2A\left(  t_{0},t_{1}\right)
\widetilde{B}\left(  t,t_{1}\right)  \right]  \overrightarrow{k}-\int_{t_{0}%
}^{t}\frac{\delta\left(  t^{\prime},t_{1}\right)  }{4}dt^{\prime}%
\end{align*}

\item for the $b\longrightarrow a$ transition:%
\[
\text{{\Large S}}_{1,ab}=e^{i\Phi_{ab}}e^{\frac{i}{\hbar}\overrightarrow
{r_{op}}.\overrightarrow{p_{ab}}}e^{-\frac{i}{\hbar}\overrightarrow{p_{op}%
}.\overrightarrow{r_{ab}}}S_{ab}\left(  \overrightarrow{r_{op}}+\widetilde
{B}\left(  t_{0},t_{1}\right)  \frac{\hbar\overrightarrow{k}}{2m}%
,\overrightarrow{p_{op}}-\widetilde{A}\left(  t_{0},t_{1}\right)  \frac
{\hbar\overrightarrow{k}}{2}\right)
\]
with:
\begin{align*}
\overrightarrow{p_{ab}}  &  =-\frac{\widetilde{A}\left(  t,t_{1}\right)
+\widetilde{A}\left(  t_{0},t_{1}\right)  }{2}\hbar\overrightarrow{k}\\
\overrightarrow{r_{ab}}  &  =+\frac{\widetilde{B}\left(  t,t_{1}\right)
+\widetilde{B}\left(  t_{0},t_{1}\right)  }{2}\frac{\hbar\overrightarrow{k}%
}{m}\\
\Phi_{ab}  &  =+\left[  \omega\frac{t+t_{0}}{2}-\omega_{0}\left(
\frac{t+t_{0}}{2}-t_{1}\right)  +\phi-\overrightarrow{k}.\frac{\overrightarrow
{\xi}\left(  t,t_{1}\right)  +\overrightarrow{\xi}\left(  t_{0},t_{1}\right)
}{2}\right] \\
&  +\frac{\hbar\overrightarrow{k}}{8m}\left[  A\left(  t,t_{1}\right)
\widetilde{B}\left(  t,t_{1}\right)  +A\left(  t_{0},t_{1}\right)
\widetilde{B}\left(  t_{0},t_{1}\right)  +2A\left(  t_{0},t_{1}\right)
\widetilde{B}\left(  t,t_{1}\right)  \right]  \overrightarrow{k}-\int_{t_{0}%
}^{t}\frac{\delta\left(  t^{\prime},t_{1}\right)  }{4}dt^{\prime}%
\end{align*}

\item for the $b\longrightarrow b$ transition:%
\[
\text{{\Large S}}_{1,bb}=e^{i\Phi_{bb}}e^{\frac{i}{\hbar}\overrightarrow
{r_{op}}.\overrightarrow{p_{bb}}}e^{-\frac{i}{\hbar}\overrightarrow{p_{op}%
}.\overrightarrow{r_{bb}}}S_{bb}\left(  \overrightarrow{r_{op}}+\widetilde
{B}\left(  t_{0},t_{1}\right)  \frac{\hbar\overrightarrow{k}}{2m}%
,\overrightarrow{p_{op}}-\widetilde{A}\left(  t_{0},t_{1}\right)  \frac
{\hbar\overrightarrow{k}}{2}\right)
\]
o\`{u}:
\begin{align*}
\overrightarrow{p_{bb}}  &  =+\frac{\widetilde{A}\left(  t,t_{1}\right)
-\widetilde{A}\left(  t_{0},t_{1}\right)  }{2}\hbar\overrightarrow{k}\\
\overrightarrow{r_{bb}}  &  =-\frac{\widetilde{B}\left(  t,t_{1}\right)
-\widetilde{B}\left(  t_{0},t_{1}\right)  }{2}\frac{\hbar\overrightarrow{k}%
}{m}\\
\Phi_{bb}  &  =-\frac{1}{2}\left[  \left(  \omega-\omega_{0}\right)  \left(
t-t_{0}\right)  -\overrightarrow{k}.\left(  \overrightarrow{\xi}\left(
t,t_{1}\right)  -\overrightarrow{\xi}\left(  t_{0},t_{1}\right)  \right)
\right] \\
&  +\frac{\hbar\overrightarrow{k}}{8m}\left[  A\left(  t,t_{1}\right)
\widetilde{B}\left(  t,t_{1}\right)  +A\left(  t_{0},t_{1}\right)
\widetilde{B}\left(  t_{0},t_{1}\right)  -2A\left(  t_{0},t_{1}\right)
\widetilde{B}\left(  t,t_{1}\right)  \right]  \overrightarrow{k}-\int_{t_{0}%
}^{t}\frac{\delta\left(  t^{\prime},t_{1}\right)  }{4}dt^{\prime}%
\end{align*}

\end{enumerate}

\begin{figure}
[h]
\begin{center}
\includegraphics[
trim=0.000000in 0.000000in -0.000627in 0.003758in,
width=362.125pt
]%
{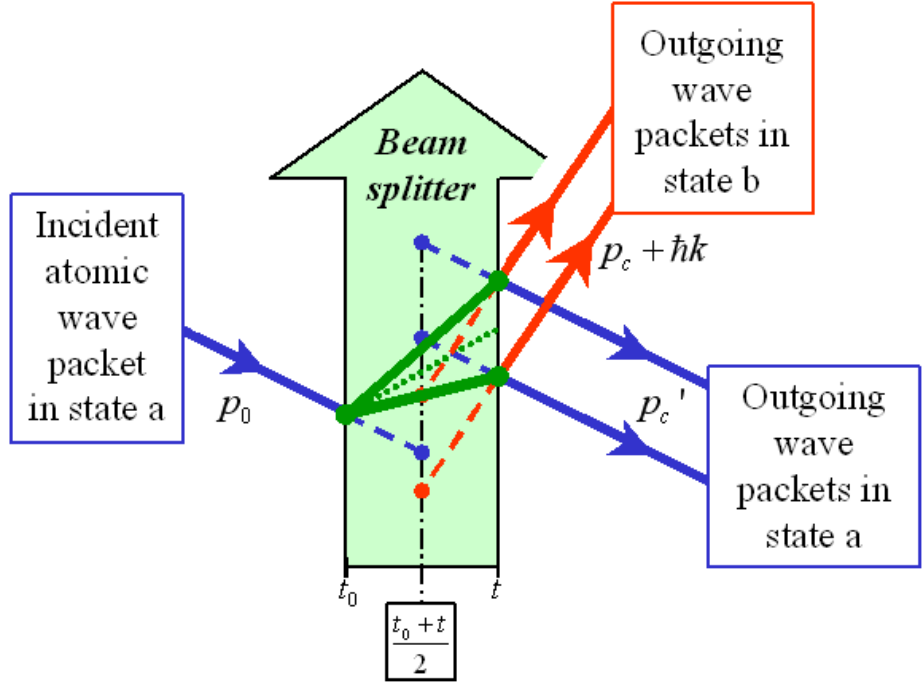}%
\caption{Atomic beam splitting in the free case and for an inelasticity
parameter different from zero. The dashed lines refer to the ttt scheme,
whereas the dotted line represents the common atomic trajectory of the
Borrmann effect.}%
\label{figttt}%
\end{center}
\end{figure}

The interpretation of these terms is simple and constitutes the core of the
generalized $ttt$ scheme:

\begin{enumerate}
\item from $t_{0}$ to $t_{1}$, the initial ket $\left\vert \Psi\left(
t_{0}\right)  \right\rangle $ evolves with $H_{0}+H_{ext}$ only (as if there
was no splitting potential);

\item at $t_{1}$, this evolved ket is subject to an effective instantaneous
interaction which modifies its structure (structuring into several wave
packets, see part \ref{part6} and the illustration of the ttt scheme on figure
\ref{figttt}), its central position and momentum, and its phase;

\item from $t_{1}$ to $t$, the obtained wave packets evolve with
$H_{0}+H_{ext}$ only, as before $t_{1}$.
\end{enumerate}

Eventually, we can show that the $t_{1}$ value which simplifies the most the
previous expressions is the central time of interaction:%

\[
t_{1}=t_{1/2}=\frac{t+t_{0}}{2}%
\]
%

Let us see now what are the features of the wave packets which appear inside
the beam splitters. Before studying this wave packet structuring in the
general case (i.e. in the presence of trapping and gravito-inertial
potentials), we will focus on the simple ``free'' case (i.e. without any such
other external potential).

\bigskip

\section{Structure of the solutions in the free case\label{part6}}

\bigskip

The study of the free case is important for at least two reasons. First, as we
have seen in part \ref{part3}, the effect of the quadratic terms of $H_{ext}$
(rotations, gradients, trap\ldots) can often be neglected during the
resolution of equation (\ref{eq3}), and any uniform acceleration can be easily
removed by modulating the laser frequency. In this case, the equation
(\ref{eq3}) is equivalent to the one stemming from the free case. Second, and
as far as we know, to date, there has been no global study of matter wave beam
splitters even in the simplest free case. Indeed, each of the previously
quoted works aims at studying only a few aspects of the atomic beam splitting:
mechanical effect, internal and external splitting, group velocities\ldots
\ Actually, the question consisting in finding what exactly goes out of an
atomic beam splitter is still relevant, even for a simple laser potential
(temporal square amplitude) and an incident Gaussian matter wave packet.

It is therefore necessary to specify what is the exact structuring of this
incident wave packet, namely the number of created wave packets (two main ones
for each transition amplitude), their amplitude (Rabi oscillations, velocity
selection, anomalous dispersion\ldots), their group velocity (Borrmann fan),
their central momentum and their phase, and eventually how these quantities
evolve in time.

\bigskip

\subsection{Group velocities and adiabatic states\label{part61}}

\bigskip

The laser-atom interaction induces particular states, the adiabatic or dressed
states, which are nothing but the eigenstates of the interaction. If
$\overline{F}\left(  t,t_{1}\right)  $ is a unit amplitude square pulse
between $t_{0}$ and $t$, these adiabatic states are simply the eigenvectors of
the matrix $iM\left(  1/2,t\right)  $. Therefore, the two corresponding
eigenenergies are:%

\[
\mp\varepsilon\left(  \overrightarrow{p}\right)  =\mp\hbar\sqrt{\Omega_{0}%
^{2}+\left[  \Delta_{l}\left(  \overrightarrow{p}\right)  /2\right]  ^{2}}%
\]

with $\Delta_{l}\left(  \overrightarrow{p}\right)  =\omega-\omega
_{0}-\overrightarrow{k}.\overrightarrow{p}/m$, and the corresponding group
velocities are simply obtained by deriving these dispersion relations with
respect to the momentum:%

\[
\pm\overrightarrow{v_{g}}\left(  \overrightarrow{p}\right)  =\mp\frac{y\left(
\overrightarrow{p}\right)  }{\sqrt{1+y\left(  \overrightarrow{p}\right)  ^{2}%
}}\frac{\hbar\overrightarrow{k}}{2m}%
\]

where $y\left(  \overrightarrow{p}\right)  $ is the well known ``off Braggness
parameter'' introduced in neutron optics \cite{rauch78,rauch00} and defined
here as:%

\[
y\left(  \overrightarrow{p}\right)  =\left(  \omega-\omega_{0}-\overrightarrow
{k}.\overrightarrow{p}/m\right)  /2\Omega_{0}%
\]

As we can see on (\ref{eq5}) and (\ref{eq6}), the matrices $S_{uv,op}$ act on
a wave packet which is shifted by a global momentum of $+\hbar\overrightarrow
{k}/2$ or $-\hbar\overrightarrow{k}/2$, depending on the initial internal
state of atoms. For the example previously described (atoms initially in the
lower internal state), we obtain the following parameter:%

\begin{equation}
y\left(  \overrightarrow{p}+\hbar\overrightarrow{k}/2\right)  =y_{+}\left(
\overrightarrow{p}\right)  =\left(  \omega-\omega_{0}-\overrightarrow
{k}.\overrightarrow{p}/m-\delta\right)  /2\Omega_{0} \label{eqy+}%
\end{equation}

which can be called ``inelasticity parameter'' as it refers to the way the
resonance condition is fulfilled \cite{antoineThese}. In the initial
representation, these group velocities become:%

\[
\frac{\overrightarrow{p}}{m}+\frac{\hbar\overrightarrow{k}}{2m}\left(
1\pm\frac{y_{+}\left(  \overrightarrow{p}\right)  }{\sqrt{1+y_{+}\left(
\overrightarrow{p}\right)  ^{2}}}\right)
\]

The examination of these velocities leads to several important conclusions.

First, the difference between momentum and group velocity in the presence of
an electromagnetic field is naturally confirmed. For a weak inelasticity
parameter $\left\vert y_{+}\left(  \overrightarrow{p}\right)  \right\vert
\ll1$, we obtain only one group velocity for both the adiabatic states (atomic
Borrmann effect, see part \ref{part52}):%

\[
\frac{\overrightarrow{p}}{m}+\frac{\hbar\overrightarrow{k}}{2m}%
\]

whereas for $\left|  y_{+}\left(  \overrightarrow{p}\right)  \right|  \gg1$,
we obtain the two extreme velocities (defining the Borrmann fan):%

\[
\frac{\overrightarrow{p}}{m}\text{ \ and \ }\frac{\overrightarrow{p}}{m}%
+\frac{\hbar\overrightarrow{k}}{m}%
\]

But in this case, the beam splitter is inefficient as we will see thereafter.

However, we can show \cite{antoineThese} that these group velocities are
closely linked to the average momenta of the considered two level system, and
that they are more precisely equal to the most probable momenta of adiabatic
states (divided by $m$).

For $y\neq0$, two distinct atomic wave packets are created in the beam
splitter, and their physical separation may be observable in certain
conditions (more than few $\mu m$ after $10^{-4}$ $s$ for an initial atomic
coherent state of $1$ $\mu m$ width).

Finally, it is noteworthy that these group velocities depend on
$\overrightarrow{p}$. Therefore, the (optical) medium where atoms evolve is
dispersive and this effect leads to the phenomenon of anomalous dispersion
(see part \ref{part64}).

\bigskip

\subsection{Rabi oscillations}

\bigskip

Let us consider the free solution of equation (\ref{eq3}). If the internal
state of initial atoms is the lower state, we obtain in the initial picture
the following lower state:%

\[
T_{aa}\left(  \overrightarrow{p_{op}},t\right)  \left|  a\left(  t_{0}\right)
\right\rangle
\]

with:%
\[
T_{aa}\left(  \overrightarrow{p},t\right)  =\exp\left[  -i\left(  \frac{E_{a}%
}{\hbar}+\left(  \overrightarrow{p}+\frac{\hbar\overrightarrow{k}}{2}\right)
^{2}/2m\hbar+\frac{\delta}{4}-\frac{\omega-\omega_{0}}{2}\right)  \left(
t-t_{0}\right)  \right]  S_{aa}\left(  \overrightarrow{p}+\frac{\hbar
\overrightarrow{k}}{2}\right)
\]
and the following upper state:%
\[
e^{-i\left(  \omega t_{0}-\overrightarrow{k}.\overrightarrow{r_{op}}%
+\phi\right)  }T_{ba}\left(  \overrightarrow{p_{op}},t\right)  \left|
a\left(  t_{0}\right)  \right\rangle
\]

with:%
\[
T_{ba}\left(  \overrightarrow{p},t\right)  =\exp\left[  -i\left(  \frac{E_{b}%
}{\hbar}+\left(  \overrightarrow{p}+\frac{\hbar\overrightarrow{k}}{2}\right)
^{2}/2m\hbar+\frac{\delta}{4}+\frac{\omega-\omega_{0}}{2}\right)  \left(
t-t_{0}\right)  \right]  S_{ba}\left(  \overrightarrow{p}+\frac{\hbar
\overrightarrow{k}}{2}\right)
\]

$S_{aa}$ and $S_{ba}$ having the following usual expressions:%

\[
S_{aa}\left(  \overrightarrow{p},t-t_{0}\right)  =\cos\left[  \varepsilon
\left(  \overrightarrow{p}\right)  .\left(  t-t_{0}\right)  \right]
-i\frac{y\left(  \overrightarrow{p}\right)  }{\sqrt{1+y\left(  \overrightarrow
{p}\right)  ^{2}}}\sin\left[  \varepsilon\left(  \overrightarrow{p}\right)
.\left(  t-t_{0}\right)  \right]
\]

\[
S_{ba}\left(  \overrightarrow{p},t-t_{0}\right)  =i\frac{1}{\sqrt{1+y\left(
\overrightarrow{p}\right)  ^{2}}}\sin\left[  \varepsilon\left(
\overrightarrow{p}\right)  .\left(  t-t_{0}\right)  \right]
\]

The temporal evolution of the initial state is therefore sinusoidal with an
amplitude equal to $1/\left(  1+y_{+}\left(  \overrightarrow{p}\right)
^{2}\right)  $ and a frequency equal to $\Omega_{0}\sqrt{1+y_{+}\left(
\overrightarrow{p}\right)  ^{2}}/2\pi$. Tuning the interaction parameters, we
can realize true atomic mirrors or \textquotedblleft$\pi$
pulses\textquotedblright\ (transfer of all the atoms from one state to
another) and semi-reflecting plates or \textquotedblleft$\pi/2$
pulses\textquotedblright\ (50-50 splitting), whose efficiency depends in a
crucial way of the value of $y_{+}\left(  \overrightarrow{p}\right)  $ which
is actually taken by the central momentum of the incident wave packet.

\bigskip

\subsection{Velocity selection}

\bigskip

This effect comes from the fact that the two pre-factors $1/\sqrt
{1+y_{+}\left(  \overrightarrow{p}\right)  ^{2}}$ and $y_{+}\left(
\overrightarrow{p}\right)  /\sqrt{1+y_{+}\left(  \overrightarrow{p}\right)
^{2}}$, in the expression of $S_{aa}$\ and $S_{ba}$, depend on the momentum
$\overrightarrow{p}$, and more particularly on its part which is collinear to
the laser wave vector $\overrightarrow{k}$ (transverse momentum). Therefore,
the terms $S_{uv}$ act as momentum filters on the incident atomic momentum distribution.

For example, in the case of the $a\longrightarrow b$ transition, one obtains
the well known sinus cardinal filter. It is characterized by a central lobe
with an amplitude of $\sin\left[  \Omega_{0}\tau\right]  $ and a total width
of $4\Omega_{0}\left(  \sqrt{\left(  \pi/\Omega_{0}\tau\right)  ^{2}%
-1}\right)  /k$ (for a fixed interaction duration of $\tau=t-t_{0}$). This
width may be less than the transverse width of the incident atomic momentum
distribution. The resulting (transverse) velocity selection is very useful in
atom interferometry (to increase the fringes contrast) and constitutes the
basis of Raman cooling \cite{kasevich92}.

\begin{figure}
[h]
\begin{center}
\includegraphics[
trim=0.000000in 0.000000in 0.001975in 0.004453in,
width=313.875pt
]%
{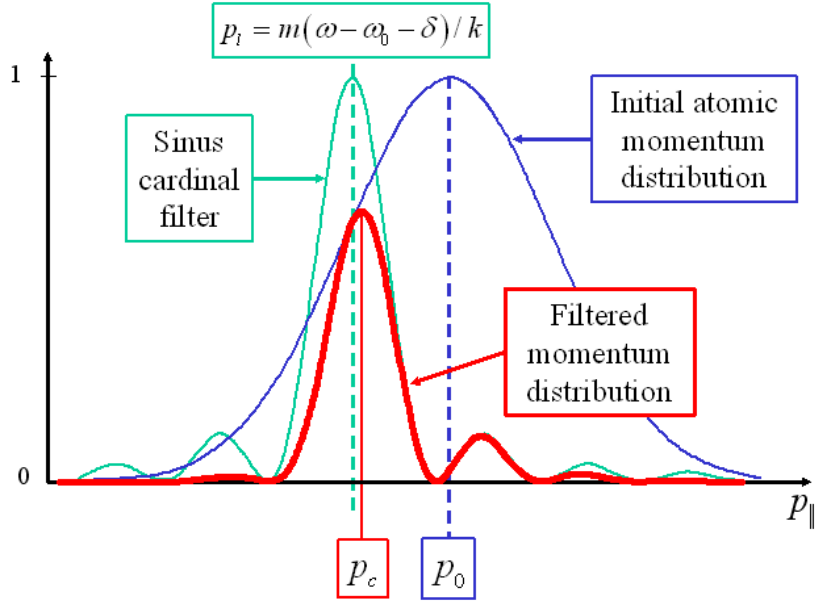}%
\caption{Transverse momentum distribution, before and after filtering by a
$\pi$ beam splitter, of a Gaussian atomic wave packet.}%
\label{figveloselect}%
\end{center}
\end{figure}

Furthermore, this central lobe is centered on the transverse momentum $p_{l}$:%

\[
p_{l}=m\left(  \omega-\omega_{0}-\delta\right)  /k
\]

which can be quite different from the initial transverse central momentum
$\overrightarrow{p_{0}}.\overrightarrow{k}/k$. If so, the central momentum
$\overrightarrow{p_{c}}$ of the filtered distribution is distinct from
$\overrightarrow{p_{0}}$ and is bounded by $\overrightarrow{p_{0}}$ and
$\overrightarrow{p_{l}}$. Consequently, the central momentum of the outgoing
wave packets (in excited state) is not $\overrightarrow{p_{0}}+\hbar
\overrightarrow{k}$, as one would think in view of the mechanical recoil
effect of light, but $\overrightarrow{p_{c}}+\hbar\overrightarrow{k}$ (see
Figure \ref{figveloselect}).%

Apart from the central lobe, the $sinc$ filter has nodes and sidelobes with an
amplitude that rapidly decreases (this particular structure is linked to the
pulse shape, and can be softened or removed by the use of apodization
functions [Blackman for example] to tailor laser pulses). If the incident
atomic distribution is sufficiently broad to encompass one or several
sidelobes, the filtered matter wave packet will then be structured into
several wave packets whith central momenta quite different from
$\overrightarrow{p_{0}}$, $\overrightarrow{p_{l}}$ or $\overrightarrow{p_{c}}$.

With respect to the other transition amplitude ($a\longrightarrow a$, i.e.
without change of internal state), we obtain the complement filter and the
structuring can be studied in the same way \cite{antoineThese}.

\bigskip

\subsection{Anomalous dispersion \label{part64}}

\bigskip

This effect refers to the modification of spreading of atomic wave packets
inside a beam splitter. It is more convenient to examine it for an incident
atomic wave packet which has a thin momentum distribution, although it can be
studied and modeled for any momentum distribution (see \cite{antoineThese}).

As we consider the wave packets which evolve inside the beam splitter, we have
to go into the adiabatic states. In the adiabatic picture, these wave packets
have the two energies (written here for the $a\longrightarrow b$ transition):%

\[
\pm\hbar\Omega_{0}\sqrt{1+y_{+}\left(  \overrightarrow{p}\right)  ^{2}}%
\]

where the term $\sqrt{1+y_{+}\left(  \overrightarrow{p}\right)  ^{2}}$ can be
Taylor expanded with respect to $\overrightarrow{k}.\left(  \overrightarrow
{p}-\overrightarrow{p_{0}}\right)  /2m\Omega_{0}$ as:%

\[
\sqrt{1+y_{+}\left(  \overrightarrow{p}\right)  ^{2}}=\sqrt{1+y_{0+}^{2}%
}-\frac{y_{0+}}{\sqrt{1+y_{0+}^{2}}}\frac{\overrightarrow{k}.\left(
\overrightarrow{p}-\overrightarrow{p_{0}}\right)  }{2m\Omega_{0}}+\frac
{1}{\left(  1+y_{0+}^{2}\right)  ^{3/2}}\frac{\left(  \overrightarrow
{p}-\overrightarrow{p_{0}}\right)  .\overset{\Rightarrow}{\delta}.\left(
\overrightarrow{p}-\overrightarrow{p_{0}}\right)  }{4m\hbar\Omega_{0}^{2}%
}+...
\]

where $y_{0+}$ is defined as:%

\[
y_{0+}=y_{+}\left(  \overrightarrow{p_{0}}\right)
\]

and where $\overset{\Rightarrow}{\delta}$ is the complement matrix of the
recoil term $\delta$:%

\[
\overset{\Rightarrow}{\delta}=\frac{\hbar\overrightarrow{k}.\widetilde
{\overrightarrow{k}}}{2m}=\frac{\hbar}{2m}\left(
\begin{array}
[c]{ccc}%
k_{x}^{2} & k_{x}k_{y} & k_{x}k_{z}\\
k_{x}k_{y} & k_{y}^{2} & k_{y}k_{z}\\
k_{x}k_{z} & k_{y}k_{z} & k_{z}^{2}%
\end{array}
\right)
\]

In the initial picture, the first order term of this expansion gives the group
velocities obtained in part \ref{part61}. As for the second order term, it
corresponds to an additional dispersion and it indicates that one wave packet
spreads more than the natural spreading, and that the other one spreads less.
In certain conditions, this spreading can even be stopped or changed into a
contraction \cite{antoineThese,eiermann03}.

In the case of a non-thin incident atomic wave packet, the main results of
this study are still valid provided $y_{0+}$ is changed into $y_{c+}%
=y_{+}\left(  \overrightarrow{p_{c}}\right)  $. One can then model the
outgoing atomic wave packets. A simple but powerful way to do this is the
Gaussian modeling which consists in writing these wave packets as Gaussians.
This ``strong field ttt modeling'', and more generally the use of Gaussian
wave packets, is found to be particularly relevant in atom interferometry (see
\cite{antoineThese}).

\bigskip

\section{Structure of the solutions in the general case\label{part7}}

\bigskip

We have already seen in part \ref{part4} how to deal with the double
non-commutation problem which appears in $\Delta_{op1}$ and in the resolution
of (\ref{eq3}). In particular, we have seen that one can often neglect, in the
expression of $\Delta_{op1}$, the terms depending on the quadratic terms of
$H_{ext}$, namely $\alpha$, $\beta-1$ and $\gamma$. In certain configurations
(when $\overrightarrow{k}$ is orthogonal to $\overrightarrow{g}$ or when
$\omega$ is modulated to compensate the gravity induced Doppler shift), the
generalized detuning is even equal to the free one and it boils down to the
free case in (and only in) the resolution of (\ref{eq3}).

In this case, we obtain the same previous adiabatic energies and consequently
the same adiabatic group velocities, Rabi oscillations, velocity selection and
anomalous dispersion effect as stated before. In the initial picture (i.e. in
the lab frame), the elements of the previous {\Large S}$_{_{1}}$ matrix are
therefore expressed as ($u,v=a,b$):%

\[
\text{{\Large S}}_{1,uv}=\exp\left[  i\Phi_{uv}\right]  .\exp\left[  \frac
{i}{\hbar}\overrightarrow{r_{op}}.\overrightarrow{p_{uv}}\right]  .\exp\left[
-\frac{i}{\hbar}\overrightarrow{p_{op}}.\overrightarrow{r_{uv}}\right]
.S_{uv}\left(  \overrightarrow{p_{op}}\mp\frac{\hbar\overrightarrow{k}}%
{2}\right)
\]

where the elements $S_{uv}$ are equal to the ones obtained in the free case
(see the previous part).

Finally, the only differences from the free case lie in the expression of:

\begin{enumerate}
\item the effective incident wave packet at time $t_{1}$ (which is equal to
the initial wave packet evolved from $t_{0}$ to $t_{1}$ thanks to
$H_{0}+H_{ext}$). In particular, its central position and momentum are no more
$\overrightarrow{r_{0}}+\overrightarrow{p_{0}}\left(  t_{1}-t_{0}\right)  /m$
and $\overrightarrow{p_{0}}$ but $\overrightarrow{R}\left(  t_{1}%
,t_{0},\overrightarrow{r_{0}},\overrightarrow{p_{0}}\right)  $ and
$\overrightarrow{P}\left(  t_{1},t_{0},\overrightarrow{r_{0}},\overrightarrow
{p_{0}}\right)  $ (expressed with the $ABCD$\ matrices, see part \ref{part3})

\item the terms $\overrightarrow{r_{uv}}$, $\overrightarrow{p_{uv}}$ and
$\Phi_{uv}$ which are detailed in part \ref{part53}

\item $U_{1}\left(  t,t_{1}\right)  $ which accounts for the evolution from
$t_{1}$ to $t$ due to $H_{0}+H_{ext}$.
\end{enumerate}

Then, we can do the same Gaussian modeling as in the free case previously
studied (see \cite{antoineThese}).

\bigskip

In some cases however, we can not neglect the effect of (unavoidable)
gravitational, inertial and trapping potentials for the resolution of
(\ref{eq3}), and it is important to examine what are the changes, due to
$H_{ext}$, of the properties listed before (group velocities, Rabi
oscillations, velocity selection and anomalous dispersion). Moreover, we have
already underlined that the key parameter is the inelasticity parameter $y$
which is proportional to the generalized detuning $\Delta$. This $y$ parameter
generally depends on the two canonical operators $\overrightarrow{r_{op}}$ and
$\overrightarrow{p_{op}}$, but in the case of linear potentials (uniform
acceleration for example) it depends on one operator ($\overrightarrow{p_{op}%
}$) only.

Before dealing with the general case, let us consider the effect of a uniform
(but time-dependent) acceleration $\overrightarrow{g}\left(  t\right)  $. From
the equation (\ref{eq3}), which becomes scalar in momentum representation, one
can extract the adiabatic energies and then the adiabatic group velocities
(written here for $t_{1}=t_{0}$):%

\[
\pm\overrightarrow{v_{g0}}\left(  t,t_{0}\right)  =\mp\frac{y_{0}}%
{\sqrt{1+y_{0}^{2}}}\frac{\hbar\overrightarrow{k}}{2m}%
\]

with ($\overline{F}$ is taken constant and equal to $1$ for simplicity):%

\[
y_{0}=y\left(  \overrightarrow{p_{0}}\right)  =y_{0free}-\overrightarrow
{k}.\int_{t_{0}}^{t}\overrightarrow{g}\left(  t^{\prime}\right)  dt^{\prime
}/2\Omega_{0}%
\]

where $y_{0free}$ is the inelasticity parameter obtained without
$\overrightarrow{g}$ (free case, see (\ref{eqy+})).

In the initial picture, this result becomes eventually:%

\[
\frac{\overrightarrow{p_{0}}}{m}+\frac{\hbar\overrightarrow{k}}{2m}%
+\int_{t_{0}}^{t}\overrightarrow{g}\left(  t^{\prime}\right)  dt^{\prime}%
\pm\overrightarrow{v_{g0}}%
\]

where two sources of atomic trajectories bending can be identified: the common
gravitational bending coming from the third term $\int_{t_{0}}^{t}%
\overrightarrow{g}\left(  t^{\prime}\right)  dt^{\prime}$, and an
\textquotedblleft anomalous gravity induced Doppler bending\textquotedblright%
\ coming from the time-dependent adiabatic group velocities. As for the free
case, this anomalous bending can be described either in terms of effective
mass tensors \cite{lammerzahl99,eiermann03} or in terms of position (and time)
dependent effective refractive index inside the beam splitter
\cite{eiermann03,agrawal01}.

\bigskip

\begin{figure}
[h]
\begin{center}
\includegraphics[
trim=0.000000in 0.000000in -0.005011in 0.006313in,
width=197.5625pt
]%
{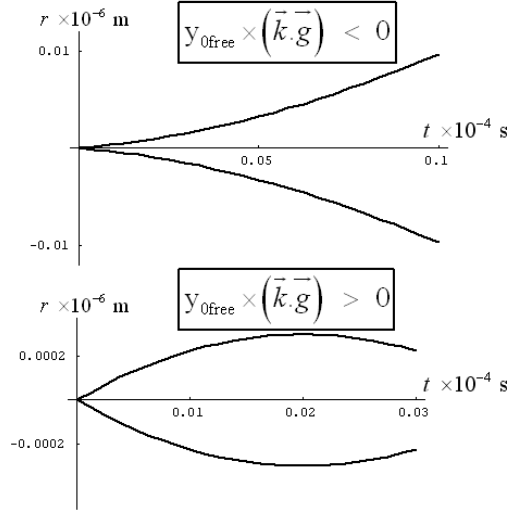}%
\caption{Central atomic trajectories inside a (temporal) beam splitter in the
presence of Earth gravity: adiabatic picture. Numerical application:
$k=10^{7}m^{-1}$, $g=10m.s^{-2}$, $\Omega_{0}=3.10^{3}s^{-1}$, $y_{0free}%
=\pm0.1$, $t_{0}=0s$. The central atomic trajectories either attract or repel
each other.}%
\label{figkg}%
\end{center}
\end{figure}

Let us remark that the acceleration induced Doppler term $-\overrightarrow
{k}.\overrightarrow{g}$ may produce a non-trivial bending of central atomic
trajectories. Indeed, if $y_{0free}$ and the scalar $\overrightarrow
{k}.\overrightarrow{g}$ are non zero, and if $y_{0free}$ has the same sign as
$\overrightarrow{k}.\overrightarrow{g}$, then the trajectories bending due to
the $\overrightarrow{k}.\overrightarrow{g}$ term makes these trajectories get
closer (as usually considered in literature). On the contrary, if $y_{0free}$
has the opposite sign of $\overrightarrow{k}.\overrightarrow{g}$, then these
two main trajectories repel each other (see Figure \ref{figkg}). In both
cases, if the vectors $\overrightarrow{k}$ and $\overrightarrow{g}$ point at
the same direction ($\overrightarrow{k}.\overrightarrow{g}>0$), downward for
example, then one of the two main atomic wave packets will always be
accelerated upward, even if all \textquotedblleft forces\textquotedblright%
\ (acceleration and laser pulse) seem to push the atoms downward.%

\begin{figure}
[h]
\begin{center}
\includegraphics[
trim=0.000000in 0.000000in -0.000532in 0.004411in,
width=301.5pt
]%
{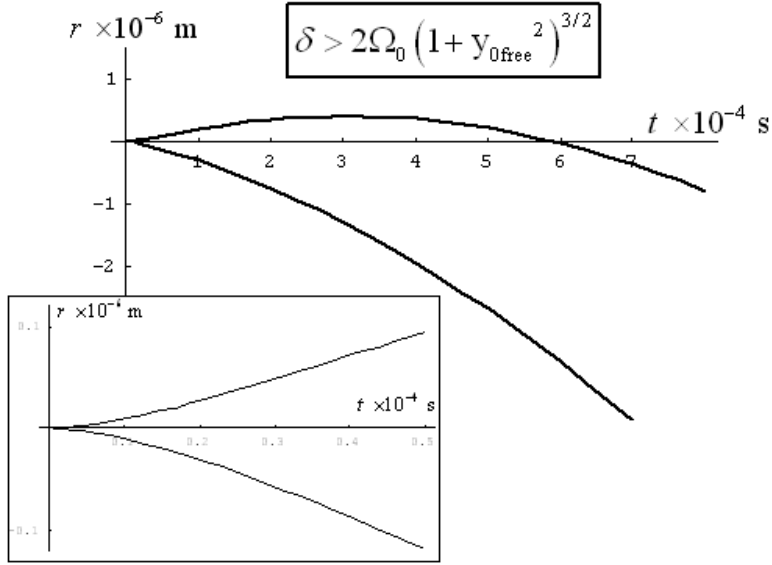}%
\caption{Central atomic trajectories inside a (temporal) beam splitter in the
presence of Earth gravity: initial picture (lab frame). Numerical application:
$k=10^{7}m^{-1}$, $g=10m.s^{-2}$, $\Omega_{0}=3.10^{3}s^{-1}$, $y_{0free}%
=-0.1$, $t_{0}=0s$. For short times, the anomalous upward acceleration exceeds
the downward acceleration of gravity.}%
\label{figantigrav}%
\end{center}
\end{figure}

Furthermore, this particular behavior is not limited to the adiabatic picture
and can be observed fully in the laboratory frame (i.e. in the initial
picture). Indeed, if $\delta>2\Omega_{0}\left(  1+y_{0free}^{2}\right)
^{3/2}$ (which corresponds, for $y_{0free}=0$, to the \textquotedblleft Bragg
regime\textquotedblright, as it is defined in \cite{oberthaler99}\ in contrast
with the \textquotedblleft channeling regime\textquotedblright\ for which
$\delta<2\Omega_{0}$), this anomalous upward acceleration exceeds the downward
acceleration of gravity (for $t-t_{0}$ much less than $1/\Omega_{0}$). As a
result, some atoms are accelerated upward in the lab frame, even if the action
of $\overrightarrow{g}$\ and laser beams is directed downward (see Figure
\ref{figantigrav}). Of course, this anomalous \textquotedblleft
anti-gravitational\textquotedblright\ bending can be explained simply by
considering the conservation of momentum and energy during the interaction
process.%

\bigskip

\bigskip

In the general case of an at-most-quadratic $H_{ext}$ (with quadratic terms
like rotations, gradients of acceleration, trapping potentials, etc), the
generalized detuning $\Delta_{op1}$ depends on the two non-commuting canonical
operators, and it can not be made scalar in any representation. As we saw
previously, one of the most relevant strategies is to approximate, in the
resolution of equation (\ref{eq3}), the effect of one of these two operators
by considering its action on a typical atomic wave packet which evolves inside
the beam splitter. In the free case, the Gaussian approximation of these
typical wave packets, called here \textquotedblleft wp\textquotedblright\ for
convenience, leads to (in the adiabatic picture):%

\begin{equation}
\overrightarrow{r_{op}}\left(  wp\right)  \simeq\left(  \overrightarrow{r_{0}%
}\pm\int_{t_{0}}^{t}\overrightarrow{v_{g0}}\left(  t^{\prime},t_{0}\right)
dt^{\prime}+\left(  U_{0}\mp U_{AD}\right)  .\left(  \overrightarrow
{p}-\overrightarrow{p_{0}}\right)  \right)  .wp \label{eq7}%
\end{equation}

where $U_{0}$ is linked to the complex momentum width of the initial atomic
wave packet, and where $U_{AD}$ accounts for the anomalous dispersion phenomenon.

In fact, we can make $U_{AD}$ arbitrarily small (for a sufficiently small
interaction duration) and incorporate the time-independent matrix $U_{0}$ into
the expression of $y$\ (see (\ref{eqdeltaop}) and (\ref{eqy+})), and consider
only the two first terms of (\ref{eq7}). Finally, the quantity
$\overrightarrow{x}\left(  t\right)  =\int_{t_{0}}^{t}\overrightarrow{v_{g0}%
}\left(  t^{\prime},t_{0}\right)  dt^{\prime}$ meets the following first order
differential equation (which can be solved numerically):%

\[
\frac{d}{dt}\overrightarrow{x}=-\frac{\overline{y_{0}}-\overrightarrow
{k}.\overset{\cdot}{A}\left(  t,t_{0}\right)  .\overrightarrow{x}/2\Omega
_{0}\overline{F}\left(  t\right)  }{\sqrt{1+\left(  \overline{y_{0}%
}-\overrightarrow{k}.\overset{\cdot}{A}\left(  t,t_{0}\right)
.\overrightarrow{x}/2\Omega_{0}\overline{F}\left(  t\right)  \right)  ^{2}}%
}\overset{\cdot}{\widetilde{B\left(  t,t_{0}\right)  }}\frac{\hbar
\overrightarrow{k}}{2m}%
\]

with:%

\[
\overline{y_{0}}=\left(  \omega-\omega_{0}-\delta-\overrightarrow{k}%
.\overset{\cdot}{A}\left(  t,t_{0}\right)  .\overrightarrow{r_{0}%
}-\overrightarrow{k}.\overset{\cdot}{B}\left(  t,t_{0}\right)
.\overrightarrow{p_{0}}/m-\overrightarrow{k}.\overset{\cdot}{\overrightarrow
{\xi}}\left(  t,t_{0}\right)  \right)  /2\Omega_{0}\overline{F}\left(
t\right)
\]

\bigskip

A simpler approximation (WKB approximation) amounts to neglecting also
$\int_{t_{0}}^{t}\overrightarrow{v_{g0}}\left(  t^{\prime},t_{0}\right)
dt^{\prime}$ in the expression (\ref{eq7}). In this case, we obtain readily:%

\[
\pm\overrightarrow{v_{g0}}\left(  t,t_{0}\right)  =\mp\frac{\overline{y_{0}}%
}{\sqrt{1+\overline{y_{0}}^{2}}}\overset{\cdot}{\widetilde{B\left(
t,t_{0}\right)  }}\frac{\hbar\overrightarrow{k}}{2m}%
\]

which gives a good approximation of the group velocities inside a matter wave
beam splitter when this latter is subject to the various external potentials
described by $H_{ext}$.

For example, for a non-uniform (but time-independent) acceleration $\left(
\overrightarrow{g},\gamma\right)  $ (or in the case of a trapping potential,
for which the sign of $\gamma$ has to be reversed), one obtains:%

\begin{align*}
\overline{y_{0}}  &  =\left(  \omega-\omega_{0}-\delta-\overrightarrow
{k}.\cosh\left(  \gamma^{1/2}\left(  t-t_{0}\right)  \right)  .\overrightarrow
{p_{0}}/m\right. \\
&  \left.  -\overrightarrow{k}.\gamma^{1/2}\sinh\left(  \gamma^{1/2}\left(
t-t_{0}\right)  \right)  .\overrightarrow{r_{0}}-\overrightarrow{k}%
.\gamma^{-1/2}\sinh\left(  \gamma^{1/2}\left(  t-t_{0}\right)  \right)
.\overrightarrow{g}\right)  /2\Omega_{0}\overline{F}\left(  t\right)
\end{align*}

Similarly, in the case of a (time-independent) rotation $\overrightarrow
{\Omega}$, one obtains:%

\[
\overline{y_{0}}=\left(  \omega-\omega_{0}-\delta-\overrightarrow
{k}.\mathcal{R}\left(  t,t_{0}\right)  .\left[  \overrightarrow{p_{0}%
}/m-\overrightarrow{\Omega}\times\left(  \overrightarrow{r_{0}}%
+\overrightarrow{p_{0}}\left(  t-t_{0}\right)  /m\right)  \right]  \right)
/2\Omega_{0}\overline{F}\left(  t\right)
\]
where the rotation matrix $\mathcal{R}\left(  t,t_{0}\right)  $ can be seen as
acting either on the atomic wave packet or on the wave vector $\overrightarrow
{k}$.

\bigskip

\section{Conclusion}

\bigskip

In conclusion, we have shown in this paper how to solve the problem of matter
wave splitting in the presence of various gravitational, inertial and trapping
potentials. In particular, we have seen how the resonance condition between
the splitting potential and the effective two-level atoms has to be changed.
Then, we have shown how to express this triple interaction \textquotedblleft
matter - splitting potential - other external potentials\textquotedblright\ as
an equivalent instantaneous interaction (generalized $ttt$ scheme).

Finally, we have investigated in detail what is the dispersive structuring of
an incident atomic wave packet inside such beam splitters, both in the free
case (for which $H_{ext}$\ is reduced to $\overrightarrow{p}^{2}/2m$) and in
the general case of $H_{ext}$. Several significant features of the solutions
have been studied: group velocities, generalized Rabi oscillations, velocity
selection, anomalous dispersion effects... In the light of this study, the
generalized $ttt$ scheme leads to a very practical and efficient (Gaussian)
modeling of atomic beam splitters which is particularly relevant for atom
interferometric signal calculations \cite{antoinejopb,antoineThese}.

It is worth pointing out that these results stem from a strong field theory
for both the splitting potential and the other external potentials described
by $H_{ext}$.

\bigskip

However, several points still have to be cleared up: for instance, the problem
of the effective mass change which occurs when the atomic internal state is
changed, and which leads to non-trivial (small) relativistic corrections.
Furthermore, it is necessary to extend our formalism to additional external
potentials which are more than quadratic (in position and momentum) if we want
to investigate the effect of van-der-Waals, Casimir or Yukawa-type potentials
on the matter wave splitting. More generally speaking, it would be interesting
to go beyond the various approximations listed in part \ref{part2}, and in
particular beyond the two-beam approximation.

\end{document}